\documentclass[a4paper,fleqn,usenatbib]{mnras}
\pdfoutput=1
\usepackage{newtxtext,newtxmath}
\usepackage[T1]{fontenc}
\usepackage{ae,aecompl}

\usepackage{graphicx}	% Including figure files
\usepackage{amsmath}	% Advanced maths commands
\usepackage{amssymb}	% Extra maths symbols

\usepackage{lscape}

\usepackage{siunitx}

\title[Rapid grain growth in post-AGB discs]{Rapid grain growth in post-AGB disc systems from far-infrared and sub-millimetre photometry\thanks{Scripts to reproduce this analysis are available at \url{http://github.com/pscicluna/GrainGrowthInPostAGBs}. They are also available on figshare along with the data at \url{https://figshare.com/account/home\#/projects/75495} } }

\author[P. Scicluna et al.]{
P. Scicluna,$^{1}$\thanks{E-mail: peterscicluna@asiaa.sinica.edu.tw}
F. Kemper$^{2,1}$, A. Trejo$^{1}$, J.~P. Marshall$^{1}$, S. Ertel$^{3}$, M. Hillen$^{4}$, %,\newauthor H. van Winckel$^{3}$ %anyone else?
\\
$^{1}$Academia Sinica, Institute of Astronomy and Astrophysics, 11F Astronomy-Mathematics Building, NTU/AS campus, \\ No. 1, Section 4, Roosevelt Rd., Taipei 10617, Taiwan\\%, Republic of China\\
$^{2}$European Southern Observatory, Karl-Schwarzschild-Str. 2, 85748 Garching b. M\"unchen, Germany\\
$^{3}$Steward Observatory, Department of Astronomy, University of Arizona, 993 N. Cherry Ave, Tucson, AZ 85721, USA\\
$^{4}$Instituut voor Sterrenkunde, KU Leuven, Celestijnenlaan 200B, 3001, Leuven, Belgium
}

\date{Accepted XXX. Received YYY; in original form ZZZ}

\pubyear{2018}

\begin{document}
\label{firstpage}
\pagerange{\pageref{firstpage}--\pageref{lastpage}}
\maketitle

% Abstract of the paper
\begin{abstract}
% This is a simple template for authors to write new MNRAS papers.
% The abstract should briefly describe the aims, methods, and main results of the paper.
% It should be a single paragraph not more than 250 words (200 words for Letters).
% No references should appear in the abstract.
{ The timescales on which astronomical dust grows remain poorly understood, with important consequences for our understanding of processes like circumstellar disk evolution and planet formation. }
A number of post-asymptotic giant branch stars are found to host optically thick, dust- and gas-rich circumstellar discs in Keplerian orbits. 
These discs exhibit evidence of dust evolution, similar to protoplanetary discs; however since post-AGB discs have substantially shorter lifetimes than protoplanetary discs they may provide new insights on the grain-growth process.
We examine a sample of post-AGB { stars} with discs to determine the FIR and sub-mm spectral index by homogeneously fitting a sample of data from \textit{Herschel}, the SMA and the literature.
We find that grain growth to at least hundreds of micrometres is ubiquitous in these systems, and that the distribution of spectral indices is more similar to that of protoplanetary discs than debris discs. 
No correlation is found with the mid-infrared colours of the discs, implying that grain growth occurs independently of the disc structure { in post-AGB discs}. 
%Furthermore, there is no correlation with the effective temperatures of the post-AGB stars, which we take as a proxy of the post-AGB age of the system.
%{ Given the diverse lifetimes of these }
%This implies
{ We infer} that grain growth to $\sim$mm sizes must occur on timescales {  $<<10^{5}$ yr, perhaps by orders of magnitude, } as the lifetimes of these {discs are} expected to be $\lesssim10^{5}$~yr and all objects have converged to the same state. %$\leq10^{3}$--$10^{4}$~yr, as the lifetimes of these {discs are} expected to be $<10^{5}$~yr {and all objects have converged to the same state}.
This growth timescale is short compared to the results of models for protoplanetary discs including fragmentation, and may provide new constraints on the physics of grain growth.
\end{abstract}

% Select between one and six entries from the list of approved keywords.
% Don't make up new ones.

%%%%%%%%%%%%%%%%%%%%%%%%%%%%%%%%%%%%%%%%%%%%%%%%%%

%%%%%%%%%%%%%%%%% BODY OF PAPER %%%%%%%%%%%%%%%%%%

\section{Introduction}
%Add intoductory text about grain growth here?
{ 
Understanding grain growth in a variety of environments remains an important outstanding challenge in understanding interstellar dust. 
Grains are inferred to evolve in a variety environments, such as the ISM, molecular clouds, circumstellar environments and AGN tori \citep[e.g.][]{Ossenkopf1993,Hirashita2014,Juhasz2010}.
These different environments correspond to very different regimes and mechanisms of dust processing, making them sensitive to different aspects of the physics. 
In planet formation, the growth of dust grains to millimetre sizes plays an important role, but the physics, primarily the timescale, of growth to these sizes is an area of open debate.
A wide variety of observations show evidence for grain growth at the later stages of planet formation \citep[e.g.][]{Ricci2010}, and the current generation of models tend to successfully explain observations of grain growth in more evolved YSOs \citep[e.g.][]{2010A&A...513A..79B,Birnstiel2014}.
Some observations of the earlier stages of planet formation have also suggested grain growth, in some cases in tension with the models \citep[e.g.][]{Miotello2014}.
However, these are not the only astrophysical objects where large grains are observed, suggesting that complementary studies of other classes of objects may provide new insights on the process; if grains grow on similar or shorter timescales under other comparable conditions this would provide new constraints on the physics.
}

The final stage of the evolution of { single low- and} intermediate mass stars ($M\sim 1-8 M_\odot$) is heralded by the ejection of their envelope as they evolve off the Asymptotic Giant Branch (AGB), exposing the core of the star as a white dwarf, ionising the ejecta and producing a spectacular planetary nebula (PN).
In particular, this phase is characterised by the development of strong asymmetries in the circumstellar medium. Many AGB envelopes are broadly spherical in shape, whereas a large fraction of PNe and pre-PNe host jets, tori, rings, or bipolar structures \citep[e.g.][]{sahai2007}.
These asymmetries are, therefore, believed to develop either in the final phase of AGB evolution or in the initial post-AGB phase.

In { many} cases, post-AGB stars are found to have a spectral energy distribution (SED) with a stronger near-infrared (NIR) excess than would be expected for an expanding spherical shell; by analogy with protoplanetary discs (PPDs) it is assumed that these are host to a massive, dusty circumstellar disc that keeps dust in the hot, inner regions of the system \citep[e.g.][]{2003ARA&A..41..391V, deruyter2005, deruyter2006, 2014A&A...568A..12H, Gezer2015,Kamath2014,Kamath2015}.
For a small number of sources the disc nature has been confirmed by scattered-light imaging \citep{Osterbart1997,Cohen2004,2019AJ....157..110E} or by observing gas in keplerian rotation in the sub-mm \citep[e.g.][and further work by the same authors]{bujarrabal2005,bujarrabal2015,bujarrabal2017}.
Many of these systems have been found to be binaries, suggesting that the presence of a companion may play a role in the formation of the disc \citep{vanwinckel2006,Oomen2018}.

{ As stated above,} other types of circumstellar discs, such as protoplanetary and debris discs \citep[e.g.][]{2011WilliamsCieza,2015Wyatt,2014Matthews}, are typically populated by dust grains at least up to mm sizes, as probed by the spectral index in the far-infrared (FIR) and (sub-)mm wavelength ranges \citep[e.g.][]{Roccatagliata2009,2012Gaspar,2016Macgregor,2017Marshall}.
In protoplanetary discs (PPDs), the existence of these large dust grains is believed to be linked to grain-growth processes which can take place in such dense, long-lived { \citep[typical lifetimes are several megayears, e.g.][]{fedele2010}} discs.
There is also evidence for the presence of grains of such sizes in the discs in post-AGBs \citep{molster1999} in spite of the large difference in lifetime compared to PPDs \citep[a factor of $\sim 100$, compare e.g.][]{fedele2010,bujarrabal2017}. 
%{ [FK: Molster et al. (1999, Nature, 401, 563) already show that grain growth occurs in disk sources based on the sub-mm spectral slope, see their Fig.~2.]}
It has also been noted that the SEDs of post-AGB discs have similar characteristics to PPDs \citep[particularly Type II Herbig Ae/Be stars;][]{deruyter2006} and that { radiative-transfer} models developed for PPDs also fit post-AGBs very effectively \citep{2014A&A...568A..12H,2015A&A...578A..40H}, suggesting that the two environments may experience similar processes.
However, to date such studies have either been relatively small \citep[e.g.][]{deruyter2005}, or the samples used have combined both post-AGBs with discs with pre-PNe and other objects \citep[e.g.][]{sahai2011}, making it difficult to evaluate the prevalence of grain growth. %{ [FK: An interesting thought here is that grain growth could actually result in the formation of parent bodies, on time scales of $\sim 10^5$~yr, which subsequently could give rise to a secondary debris disk. Another reason to compare our results to debris disk studies.]}
Since post-AGB discs are expected to have very short lifetimes, if large grains are common it could have significant consequences for our understanding of grain growth in circumstellar discs, placing strong constraints on the timescale of this process. 

In this paper, we present a systematic analysis of the FIR and sub-mm emission of post-AGB stars with discs and the relationship to grain growth in these objects. 
We exploit archival \textit{Herschel}/SPIRE photometry along with new SMA observations and literature fluxes of a sample of 45 post-AGBs with discs to derive the spectral index and hence infer whether significant grain growth may have occurred.
Based on this, we explore how the unique environments of these systems can constrain the physics of dust growth and the implications for the initial conditions of planet formation.
In Sect.~\ref{sect:observations} we describe the observational data employed, and Sect.~\ref{sect:methods} outlines the methods used to analyse the data.
In Sect.~\ref{sect:discussion} we discuss the implications of our results.
Finally, in Sect.~\ref{sect:conclusions}, we summarise our findings. 

\section{Observations}
\label{sect:observations}

\subsection{Sample selection}
Our sample is selected from the sample of RV Tauris with IR excess and confirmed binary post-AGBs of \citet{deruyter2006}.
We select all sources in that sample that were observed with \textit{Herschel}.
In addition, we add the famous post-AGB binary the Red Rectangle, which has very deep ALMA photometry from \citet{bujarrabal2013} and was also included in the \citet{deruyter2006} sample{ , although no Herschel data are available}.

\subsection{Herschel}

We used data from the \textit{Herschel} Space Observatory's \citep{2010Pilbratt}  far-infrared and sub-millimetre instruments Photodetector Array Camera and Spectrograph \citep[PACS;][]{2010Poglitsch} (70/100/160~$\mu$m) and Spectral and Photometric Imaging REceiver \citep[SPIRE][]{2010Griffin} (250/350/500~$\mu$m). { Imaging} observations of the targets were obtained from the \textit{Herschel} Science Archive\footnote{\url{http://archives.esac.esa.int/hsa/whsa/}} as either level 2.5 (mosaiced, pipeline-reduced) or level 3 (user-reduced) data products. In our analysis, we included data from the following observing programmes: KPOT\_smolinar\_1 \citep[HiGAL - the Herschel Infrared Galactic Plane Survey;][]{2010Molinari,2016Molinari}, OT2\_smolinar\_7, and OT2\_cgielen\_4.

\textit{Herschel} data analysis was carried out in the \textit{Herschel} Interactive Processing Environment \citep[HIPE;][]{2010Ott}. The data were taken as science-ready products from the archive as described above. For the PACS wavebands we used the JScanamorphos image, as it is the most reliable for retaining complex background structure through the spatial filtering that occurs in the map-making process. For aperture photometry we adopted an aperture radius of 20\arcsec\, and a sky annulus of 30--40\arcsec. Measured fluxes in each band were adjusted with the appropriate aperture correction factors from \cite{2014Balog}. For the SPIRE wavebands we used the implementation of the sextractor tool provided in HIPE to measure the fluxes. This is due to the bright and structured nature of the backgrounds at sub-millimetre wavelengths making aperture photometry inadvisable. The beam area and full-width half-maxima for each band were taken from the SPIRE observer's manual\footnote{\url{http://herschel.esac.esa.int/Docs/SPIRE/html/spire\_om.html}}. No colour corrections were applied to the fluxes. No frequency dependence was applied for the SPIRE maps (assumed $\nu^{-1}$ dependence, Rayleigh-Jeans tail is $\nu^{2}$). A calibration uncertainty of 5 per cent is assumed for both PACS and SPIRE measurements.

\subsection{Sub-Millimeter Array}
Four { of the} science targets (RV Tau, AY Lep, PS Gem and AR Pup) were observed on a single night (or track) on 2017/01/03 with 8 antennae in compact configuration, giving baselines up to 77\,m in length. 
The SMA can employ two receivers at the same time, to effectively double the bandwidth available. 
The track used this observing mode, with both receivers tuned to an LO frequency of 225\,GHz (1.3\,mm), with 6\,GHz continuum bandwidth per receiver per sideband.
However, technical problems resulted in the loss of all data for baselines involving one antenna for one spectral chunk (2\,GHz wide; channel width $\sim$ 560 kHz). 
The remainder of the baselines for the affected chunk could still be used.
%, giving an effective continuum bandwidth of 4\,GHz per receiver per sideband, giving a total effective continuum bandwidth of 16\,GHz and 7 antennae.
Therefore, the fraction of data lost did not have a big impact in the  final achieved sensitivity.
Depending on elevation constraints, and some data lost during the calibration process, the final time on source was from $\sim$ 1.5 to 2 hours, per target.
%Targets were observed or approximately 1.5 hours.
%giving continuum rms sensitivity of\,$\sim1$\,mJy\,beam$^{-1}$.
For the calibration, the SMA observed 3C273 for bandpass, Uranus and 2232+117 for flux, and five more quasars for temporal amplitude and phase gains, interleaved in a loop with the science targets.

The data was calibrated using {\sc mir}\footnote{\url{https://www.cfa.harvard.edu/sma/smaMIR/}}, following the standard procedures. %\citep{} apparently there's no citation for MIR!
The imaging was done with the {\sc aips} \citep{Wells1985} task IMAGR. 
The final {\it rms} varies across the continuum maps, and is between 0.5 and \,$\sim1$\,mJy\,beam$^{-1}$. 
Only AR Pup and RV Tau were detected as unresolved continuum sources, with { SNR} values of $\sim$ 50 and 16, respectively. 

The spectral setup was chosen to place the CO(2--1) line in the upper sideband. 
However, no line emission was detected towards any of the targets, even with re-binning up to $\sim$ 7 km s$^{-1}$ and {\it rms} values down to $\sim$ 20\,mJy.

% ~ 100 mJy for 0.7 km/s, or ~ 20 mJy for ~ 7km/s channels.
%with {\it rms} values of $\sim$ 100 mJy for 
%These observations are summarised in Tab.~\ref{tab:smaobs}.

\subsection{Literature}
To improve sampling at longer wavelengths we searched the literature for published sub-mm observations of the \textit{Herschel}-observed sample.
These fluxes are primarily taken from observations with SCUBA, but also include observations from ALMA and at 1.3 and 3\,mm from other facilities.
A full list of fluxes are given in Tab.~\ref{tab:fluxes} along with the \textit{Herschel} and SMA observations described above.

%\begin{landscape}
\begin{table*}%[]
    \caption{Complete list of fluxes, in mJy {\bf for our sample. Telescopes or instruments for the observations and their origin are noted in the references. Where not otherwise noted, fluxes are from {\it Herschel}}. }
    \centering
    \footnotesize
    \rotatebox{90}{
    \begin{tabular}{llccccccccc}\hline
    Source & IRAS PSC & 100 & 160 & 250 & 350 & 450 & 500 & 850 & 1300 & 3000  \\ \hline\hline
    TW Cam &04166+5719&--&--& 145 $\pm$ 10 & 74 $\pm$ 8 &--& 0 $\pm$ 7 & 11 $\pm$ 2.3$^{a}$ &--&--\\ %dR06
DY Ori &06034+1354& 1431 $\pm$ 72 & 571 $\pm$ 32 & 223 $\pm$ 14 & 114 $\pm$ 9 &--& 0 $\pm$ 7 &--&--&--\\ %dR06
CT Ori &06072+0953&--&--& 50 $\pm$ 8 & 0 $\pm$ 79 &--& 0$\pm$ 56 &--&--&--\\ %dR06
SU Gem &06108+2743& 819 $\pm$ 41 & 358 $\pm$ 22 & 166 $\pm$ 9 & 8 $\pm$ 6 &--& 0 $\pm$ 32 & 7.5 $\pm$ 2.5$^{a}$ &--&--\\ %dR06
%UY CMa & 136 $\pm$ 9 & 90 $\pm$ 12 & 0 $\pm$ 7 &--&--&--& 2.4 $\pm$ 2.1$^{a}$ &--&--\\
%UY CMa2 
UY CMa &06160-1701& 162 $\pm$ 11 & 87 $\pm$ 23 & 22.5 $\pm$ 6.7 &--&--&--& 2.4 $\pm$ 2.1$^{a}$ &--&--\\ %dR06
V382 Aur &06338+5333& 029 $\pm$ 5 & 49 $\pm$ 8 & 0 $\pm$ 7 &--&--&--&--&--&--\\ %dR06 (HD46703)
ST Pup &06472-3713& 362 $\pm$ 18 & 78 $\pm$ 9 & 30 $\pm$ 8 & 0 $\pm$ 13 &--&--&--&--&--\\ %dR06
V421 CMa &07140-2321& 201 $\pm$ 11 & 29 $\pm$ 9 & 31 $\pm$ 8 & 0 $\pm$ 7 &--&--&--&--&--\\ %dR06 (SAO 173329)
U Mon &07284-0940& 9120 $\pm$ 456 & 4236 $\pm$ 212 & 2009 $\pm$ 104 & 1066 $\pm$ 54 &--& 549 $\pm$ 32 & 181.6 $\pm$ 2.6$^{a}$ & 100 $\pm$ 14$^{b}$ & 15 $\pm$ 0.3$^{b}$ \\  %dR06
V390 Vel &08544-4431&--&--& 4229 $\pm$ 220 & 1863 $\pm$ 97 &--& 588 $\pm$ 34 &--&--&--\\ %dR06 (IRAS 08544-4431)
BZ Pyx &09060-2807& 242 $\pm$ 13 & 126 $\pm$ 11 & 0 $\pm$ 9 &--&--&--&--&--&--\\ %dR06 (IRAS 09060-2807)
IRAS 09144-4933 &09144-4933& 1107 $\pm$ 55 & 379 $\pm$ 31 & 0 $\pm$ 14 &--&--&--&--&--&--\\ %dR06
IRAS 09400-4733 &09400-4733& 582 $\pm$ 29 & 270 $\pm$ 17 & 112 $\pm$ 7 & 45 $\pm$ 5 &--& 0 $\pm$ 7 &--&--&--\\ %dR06
GP Cha &09538-7622& 163 $\pm$ 9 & 56 $\pm$ 9 & 0 $\pm$ 10 &--&--&--&--&--&--\\ %dR06 (IRAS 09538-7622)
AG Ant &10158-2844& 479 $\pm$ 24 & 235 $\pm$ 13 & 72 $\pm$ 8 & 54 $\pm$ 7 &--& 53 $\pm$ 9 & 8.7 $\pm$ 2.8$^{c}$ &--&--\\ %dR06 (HR 4049)
HR 4226 &10456-5712&--&--& 2158 $\pm$ 111 & 1093 $\pm$ 58 &--& 530 $\pm$ 30 &--&--&--\\ %dR06 (HD93662)
V802 Car &11000-6153& 1545 $\pm$ 77 & 983 $\pm$ 163 & 0 $\pm$ 14 &--&--&--&--&--&--\\ %dR06 (HD 95767)
GK Car &11118-5726& 169 $\pm$ 11 & 109 $\pm$ 14 & 0 $\pm$ 36 &--&--&--&--&--&--\\ %dR06
AF Crt &11472-0800& 321 $\pm$ 17 & 80 $\pm$ 10 & 29 $\pm$ 8 & 24 $\pm$ 8 &--& 0 $\pm$ 7 &--&--&--\\ %dR06 (IRAS 11472-0800)
RU Cen &12067-4508&--&--& 589 $\pm$ 31 & 319 $\pm$ 18 &--& 149 $\pm$ 12 &--&--&--\\ %dR06
SX Cen & 12185-4856&--&--& 83 $\pm$ 9 & 45 $\pm$ 8 &--& 0 $\pm$ 7 &--&--&--\\ %dR06
V1123 Cen &12222-4652& 2446 $\pm$ 122 & 928 $\pm$ 46 & 292 $\pm$ 17 & 126 $\pm$ 10 &--& 49 $\pm$ 80 &--&--&--\\ %dR06 (HD 108015)
%V1123 Cen2 & 2394 $\pm$ 11 & 929 $\pm$ 27 & 295 $\pm$ 15 & 132 $\pm$ 19 &--&--&--&--&--\\
IRAS 13258-8103 &13258-8103& 660 $\pm$ 33 & 458 $\pm$ 31 & 143 $\pm$ 8 & 51 $\pm$ 80 &--& 0 $\pm$ 7 &--&--&--\\ %dR06
EN TrA &14524-6838& 1459 $\pm$ 73 & 625 $\pm$ 33 & 266 $\pm$ 16 & 139 $\pm$ 10 &--& 75 $\pm$ 11 &--&--&--\\ %dR06
IRAS 15469-5311 &15469-5311& 3880 $\pm$ 194 & 1441 $\pm$ 75 & 0 $\pm$ 49 &--&--&--&--&--&--\\ %dR06 
IRAS 15556-5444 &15556-5444& 2987 $\pm$ 149 & 1363 $\pm$ 72 & 660 $\pm$ 35 & 319 $\pm$ 18 &--& 169 $\pm$ 013 &--&--&--\\ %dR06 
NSV 7708 &16230-3410& 254 $\pm$ 13 & 75 $\pm$ 8 & 41 $\pm$ 80 & 0 $\pm$ 11 &--&--&--&--&--\\ %dR06 (IRAS 16230-3410)
IRAS 17038-4815 &17038-4815& 1878 $\pm$ 94 & 994 $\pm$ 55 & 447 $\pm$ 25 & 268 $\pm$ 16 &--& 105 $\pm$ 11 &--&--&--\\ %dR06 
LR Sco &17243-4348& 762 $\pm$ 38 & 334 $\pm$ 29 & 86 $\pm$ 8 & 45 $\pm$ 6 &--& 60 $\pm$ 8 &--&--&--\\ %dR06
89 Her &17534+2603&--&--& 667 $\pm$ 36 & 343 $\pm$ 20 &--& 160 $\pm$ 13 & 40.9 $\pm$ 2.4$^{c}$ & 9.2 $\pm$ 0.5$^{d}$ & 2.7 $\pm$ 0.3$^{d}$ \\ %dR06
AI Sco &17530-3348& 1424 $\pm$ 71 & 631 $\pm$ 32 & 301 $\pm$ 17 & 14 $\pm$ 9 &--& 68 $\pm$ 8 &--&--&--\\ %dR06
V2053 Oph &18123+0511& 1213 $\pm$ 61 & 390 $\pm$ 22 & 107 $\pm$ 10 & 32 $\pm$ 7 &--& 0 $\pm$ 7 &--&--&--\\ %dR06 (IRAS 18123+0511)
IRAS 18158-3445 &18158-3445& 375 $\pm$ 19 & 135 $\pm$ 11 & 1328 $\pm$ 66 & 655 $\pm$ 33 &--& 314 $\pm$ 16 &--&--&--\\  %dR06
AC Her &18281+2149&--&--& 1572 $\pm$ 83 & 817 $\pm$ 44 &--& 380 $\pm$ 22 & 99.4 $\pm$ 3.8$^{a}$ & 38 $\pm$ 1$^{b}$ & 4.6 $\pm$ 0.4$^{b}$ \\ %dR06
AD Aql &18564-0814& 156 $\pm$ 12 & 93 $\pm$ 8 & 35 $\pm$ 8 & 0 $\pm$ 7 &--&--&--&--&--\\ %dR06
EP Lyr &19163+2745& 107 $\pm$ 7 & 80 $\pm$ 10 & 37 $\pm$ 8 & 26 $\pm$ 7 &--& 22 $\pm$ 7 &--&--&--\\ %dR06
BD-02 4931 &19157-0247& 823 $\pm$ 41 & 408 $\pm$ 22 & 151 $\pm$ 11 & 93 $\pm$ 9 &--& 41 $\pm$ 8 &--&--&--\\ %dR06 (IRAS 19157-0247)
QY Sge &20056+1834&--&--& 372 $\pm$ 22 & 191 $\pm$ 13 &--& 108 $\pm$ 14 & 218 $\pm$ 1.8$^{e}$ &--&--\\ %dR06
%AR Pup &--&--& 1597 $\pm$ 82 & 923 $\pm$ 52 &--& 443 $\pm$ 35 &--& 41.6 $\pm$ 2.1$^{f}$ &--\\
%AR Pup2 &--&--& 1504.2 $\pm$ 220 & 778.6 $\pm$ 140 &--& 378.6 $\pm$ 70 &--& 41.6 $\pm$ 2.1$^{f}$ &--\\
%AR Pup3 
AR Pup &08011-3627&--&--& 1610 $\pm$ 83 & 861 $\pm$ 96 &--& 429 $\pm$ 72 &--& 41.6 $\pm$ 2.1$^{f}$ &--\\  %dR06
RV Tau &04440+2605& 2414 $\pm$ 121 & 987 $\pm$ 49 & 512 $\pm$ 28 & 263 $\pm$ 18 &--& 135 $\pm$ 16 & 50.3 $\pm$ 3.6$^{a}$ & 15 $\pm$ 0.8$^{f}$ & 3.9 $\pm$ 0.2$^{b}$ \\  %dR06
%RV Tau2 & 2879 $\pm$ 11 & 1327 $\pm$ 26 & 539 $\pm$ 16 & 292 $\pm$ 15 &--& 147 $\pm$ 14 & 50.3 $\pm$ 3.6$^{a}$ & 15 $\pm$ 0.8$^{f}$ & 3.9 $\pm$ 0.2$^{a}$ \\
AY Lep &05208-2035& 680 $\pm$ 34 & 244 $\pm$ 12 & 102 $\pm$ 12 & 31 $\pm$ 10 &--& 0 $\pm$ 7 &--& 0 $\pm$ 1$^{f}$ &--\\ %dR06 (IRAS 05208-2035)
PS Gem &07008+1050& 338 $\pm$ 17 & 143 $\pm$ 12 & 75 $\pm$ 16 & 63 $\pm$ 15 &--& 24 $\pm$ 7 & 2.8 $\pm$ 1.9$^{c}$ & 0 $\pm$ 1$^{f}$ &--\\  %dR06 (HD52961)
IRAS 17233-4330 &17233-4330& 1024 $\pm$ 13 & 496 $\pm$ 30 & 278 $\pm$ 41 & 208 $\pm$ 48 &--& 132 $\pm$ 34 &--&--&--\\ %dR06
BD+03 3950 &19125+0343& 2331 $\pm$ 11 & 893 $\pm$ 27 & 313 $\pm$ 22 & 136 $\pm$ 21 &--& 84 $\pm$ 19 &--&--&--\\ %dR06 (IRAS 19125+0343)
IW Car &09256-6324&--&--& 2687 $\pm$ 30 & 1415 $\pm$ 27 &--& 685 $\pm$ 26 & 200 $\pm$ 20$^{g}$ &--&--\\ %dR06
Red Rectangle &06176-1036&--&--&--&--& 3400 $\pm$ 200$^{h}$ &--& 625 $\pm$ 35$^{h}$ &--&--\\\hline  %dR06 (HD 44179)
%\multicolumn{10}{l}{Notes:} \\
\multicolumn{10}{l}{References: $a$ SCUBA, \citet{deruyter2005}; $b$ CARMA, \citet{sahai2011}; $c$ SCUBA, \citet{deruyter2006}; $d$ PdBI, \citet{bujarrabal2007}; } \\
\multicolumn{10}{l}{$e$ SCUBA, \citet{2002MNRAS.332L..55G}; $f$ SMA, this work; $g$ ALMA, \citet{bujarrabal2017};
$h$ ALMA, \citet{bujarrabal2013}}
    \end{tabular}
    }
    
    \label{tab:fluxes}
\end{table*}
%\end{landscape}

%AG Ant e
%89 Her e
%QY Sgr g
%PS Gem  HD 52961

\section{Results}
\label{sect:methods}
%\subsection{Spectral indices}
The observing data described in the previous section was used to derive spectral indices in the FIR-sub-mm region for all sources with detections in two or more bands at wavelengths of 250\,$\mu$m or longer. 
To ensure that the results of the fitting are robust, and can directly integrate the existence of upper-limits on the fluxes in some filters, we employ the affine-invariant MCMC implementation {\it emcee} \citep{emcee}. 
Standard non-linear least-square minimisers are typically not able to handle censored data e.g. upper or lower limits. 
However, in many cases upper limits (left-censored data) on the flux provide important constraints on the spectral index, particularly for constraining non-detections at long wavelengths, and hence more advanced methods which are able to account for this are required.
The treatment of upper limits is trivial in MCMC, allowing us to treat sources with different numbers of detections uniformly.
The model being fitted is a straight forward power law, which becomes a straight line when fitted to the logarithm of the fluxes i.e. 
\begin{equation}
    \log F_\nu = \alpha\log\lambda + b
\end{equation}
where $F_\nu$ is the flux density in milli-Jansky, $\lambda$ is the observing wavelength and $\alpha$ is the spectral index. As the colour corrections for the filters in question are typically at most a few percent% %\citep{}
, we opt to work with this simple model as is rather than performing synthetic photometry.
For a datum with $F_\nu  / \sigma_{F_\nu} > 3$ our likelihood function is the familiar
\begin{equation}
    \log \mathcal{L}_i = -\frac{1}{2} \frac{\left(\log F_{\nu, obs} - \log F_{\nu, mod}\right)^2}{\delta_{F_\nu}^2} + \log\left(2\pi\delta_{F_\nu}^2\right)
\end{equation}
where $F_{\nu, obs}$ and $F_{\nu, mod}$ are the observed and model fluxes, respectively and $\delta_{F_\nu}$ is the fractional uncertainty $\sigma_{F_\nu} / F_\nu$, which corresponds to the uncertainty on $\log F_\nu$. 
For upper limits, this must be modified to correctly capture how well the limit constrains the model.
In these cases it is necessary to integrate the pdf of the upper limit, i.e.
\begin{equation}
    \log \mathcal{L}_i = \log \int^{3\sigma_{F_\nu}}_{0} \mathcal{N}\left(F_{\nu, mod}, \sigma_{F_\nu} \right) dF_\nu
\end{equation}
where $\mathcal{N}\left(F_{\nu, mod}, \sigma_{F_\nu} \right) $ is a Normal distribution with mean $F_{\nu, mod}$ and standard deviation $\sigma_{F_\nu}$.
In this case, models which do no violate or approach the upper limit have broadly similar likelihoods at that wavelength, while models which do are strongly penalised.
The final log-likelihood of a given model is then
\begin{equation}
    \log \mathcal{L} = \sum \log \mathcal{L}_i
\end{equation}
which is used as our objective function in the MCMC.
This model naturally results in a tight anticorrelation between the slope and intercept of the line, but this is only significant for the sources with the most uncertain data.

To explore the influence of the choice of wavelength coverage, the fitting was repeated for different subsets of the data.
The results of these fits are summarised in Tab.~\ref{tab:alphas}. %, and shown in detailed figures in the online material.
Upon examining the results, it is clear that the inclusion of PACS fluxes biases the results toward shallower slopes. 
This is most likely a result of temperature effects -- PACS samples wavelengths too close to the peak of the emission, and hence there is significant curvature in this region of the SED.
As a result, we take the results for fits including all wavelengths $\ge 250\,\mu$m (labelled ``No PACS'' in Tab.~\ref{tab:alphas}) as our preferred set, and use these in all further analyses.
%The scripts employed can be found at \url{}.

\begin{table*}%[]
    \caption{Spectral indices for different choices of dataset.}
    \centering
    \renewcommand{\arraystretch}{1.35}
    \begin{tabular}{lccccc}\hline
    Source & SPIRE only & Herschel & No Herschel & No PACS & All data  \\ \hline\hline
    TW Cam & -3.05 $\pm^{ 0.51}_{0.58}$ & -3.03 $\pm^{ 0.48}_{0.58}$ & -2.54 $\pm^{ 1.73}_{1.67}$ & -2.50 $\pm^{ 0.32}_{0.33}$ & -2.49 $\pm^{ 0.31}_{0.32}$ \\
DY Ori & -3.33 $\pm^{ 0.39}_{0.42}$ & -2.36 $\pm^{ 0.11}_{0.11}$ &--& -3.30 $\pm^{ 0.38}_{0.43}$ & -2.36 $\pm^{ 0.11}_{0.11}$ \\
%CT Ori & -2.44 $\pm^{ 1.67}_{1.74}$ & -2.53 $\pm^{ 1.74}_{1.71}$ &--& -2.46 $\pm^{ 1.66}_{1.72}$ & -2.53 $\pm^{ 1.70}_{1.70}$ \\
SU Gem & -2.20 $\pm^{ 0.61}_{0.62}$ & -1.81 $\pm^{ 0.14}_{0.14}$ & -2.55 $\pm^{ 1.73}_{1.64}$ & -2.19 $\pm^{ 0.61}_{0.62}$ & -1.81 $\pm^{ 0.14}_{0.14}$ \\
%UY CMa & -2.49 $\pm^{ 1.67}_{1.71}$ & -2.15 $\pm^{ 0.36}_{0.42}$ & -2.45 $\pm^{ 1.68}_{1.72}$ & -2.49 $\pm^{ 1.68}_{1.70}$ & -2.16 $\pm^{ 0.36}_{0.42}$ \\
%UY CMa2 
UY CMa& -2.56 $\pm^{ 1.73}_{1.66}$ & -1.94 $\pm^{ 0.66}_{0.69}$ & -2.46 $\pm^{ 1.69}_{1.73}$ & -3.10 $\pm^{ 1.39}_{1.28}$ & -2.19 $\pm^{ 0.44}_{0.58}$ \\
%V382 Aur & -2.52 $\pm^{ 1.72}_{1.68}$ & -0.91 $\pm^{ 0.51}_{0.65}$ &--& -2.48 $\pm^{ 1.68}_{1.72}$ & -0.92 $\pm^{ 0.52}_{0.67}$ \\
%ST Pup & -2.76 $\pm^{ 1.70}_{1.53}$ & -3.03 $\pm^{ 0.46}_{0.47}$ &--& -2.82 $\pm^{ 1.77}_{1.50}$ & -3.02 $\pm^{ 0.46}_{0.49}$ \\
%V421 CMa & -3.16 $\pm^{ 1.72}_{1.28}$ & -2.51 $\pm^{ 0.47}_{0.54}$ &--& -3.14 $\pm^{ 1.72}_{1.31}$ & -2.52 $\pm^{ 0.47}_{0.54}$ \\
U Mon & -1.88 $\pm^{ 0.25}_{0.25}$ & -1.74 $\pm^{ 0.09}_{0.10}$ & -1.98 $\pm^{ 0.05}_{0.04}$ & -1.98 $\pm^{ 0.03}_{0.04}$ & -1.93 $\pm^{ 0.03}_{0.03}$ \\
V390 Vel & -2.83 $\pm^{ 0.25}_{0.26}$ & -2.83 $\pm^{ 0.25}_{0.26}$ &--& -2.84 $\pm^{ 0.25}_{0.25}$ & -2.84 $\pm^{ 0.26}_{0.26}$ \\
%BZ Pyx & -2.52 $\pm^{ 1.76}_{1.70}$ & -2.32 $\pm^{ 0.29}_{0.32}$ &--& -2.50 $\pm^{ 1.70}_{1.70}$ & -2.30 $\pm^{ 0.28}_{0.32}$ \\
%IRAS 09144-4933 & -2.51 $\pm^{ 1.74}_{1.68}$ & -3.37 $\pm^{ 0.25}_{0.28}$ &--& -2.48 $\pm^{ 1.71}_{1.70}$ & -3.37 $\pm^{ 0.25}_{0.28}$ \\
IRAS 09400-4733 & -3.12 $\pm^{ 0.62}_{0.69}$ & -2.03 $\pm^{ 0.13}_{0.14}$ &--& -3.11 $\pm^{ 0.62}_{0.71}$ & -2.03 $\pm^{ 0.14}_{0.14}$ \\
%GP Cha & -2.59 $\pm^{ 1.73}_{1.66}$ & -2.63 $\pm^{ 0.59}_{0.70}$ &--& -2.42 $\pm^{ 1.67}_{1.75}$ & -2.60 $\pm^{ 0.57}_{0.71}$ \\
AG Ant$^{\ast}$ & -0.69 $\pm^{ 0.45}_{0.60}$ & -1.66 $\pm^{ 0.17}_{0.17}$ & -2.52 $\pm^{ 1.71}_{1.69}$ & -1.13 $\pm^{ 0.48}_{0.49}$ & -1.69 $\pm^{ 0.16}_{0.16}$ \\
HR 4226 & -2.03 $\pm^{ 0.26}_{0.25}$ & -2.03 $\pm^{ 0.25}_{0.26}$ &--& -2.03 $\pm^{ 0.25}_{0.25}$ & -2.02 $\pm^{ 0.26}_{0.25}$\\
%V802 Car & -2.47 $\pm^{ 1.70}_{1.72}$ & -3.86 $\pm^{ 0.30}_{0.34}$ &--& -2.62 $\pm^{ 1.75}_{1.66}$ & -3.84 $\pm^{ 0.29}_{0.34}$ \\
%GK Car &-2.52 $\pm^{ 1.71}_{1.68}$& -1.20 $\pm^{ 0.53}_{0.60}$ &--& -2.42 $\pm^{ 1.66}_{1.71}$ & -1.17 $\pm^{ 0.52}_{0.61}$ \\
AF Crt & -2.36 $\pm^{ 1.35}_{1.48}$ & -2.54 $\pm^{ 0.39}_{0.39}$ &--& -2.42 $\pm^{ 1.38}_{1.50}$ & -2.54 $\pm^{ 0.38}_{0.40}$ \\
RU Cen & -1.95 $\pm^{ 0.31}_{0.31}$ & -1.96 $\pm^{ 0.31}_{0.31}$ &--& -1.97 $\pm^{ 0.31}_{0.31}$ & -1.95 $\pm^{ 0.30}_{0.30}$ \\
SX Cen & -2.85 $\pm^{ 0.81}_{0.96}$ & -2.86 $\pm^{ 0.81}_{0.97}$ &--& -2.87 $\pm^{ 0.83}_{0.97}$ & -2.83 $\pm^{ 0.81}_{0.96}$ \\
V1123 Cen & -2.51 $\pm^{ 0.67}_{0.67}$ & -2.36 $\pm^{ 0.15}_{0.15}$ &--& -2.50 $\pm^{ 0.66}_{0.66}$ & -2.36 $\pm^{ 0.15}_{0.15}$\\ 
%V1123 Cen2 & -2.41 $\pm^{ 1.00}_{0.99}$ & -2.18 $\pm^{ 0.09}_{0.09}$ &--& -2.36 $\pm^{ 0.98}_{1.00}$ & -2.18 $\pm^{ 0.09}_{0.09}$ \\
IRAS 13258-8103 & -3.91 $\pm^{ 0.86}_{0.74}$ & -2.02 $\pm^{ 0.12}_{0.13}$ &--& -3.91 $\pm^{ 0.86}_{0.73}$ & -2.02 $\pm^{ 0.13}_{0.13}$ \\
EN TrA & -1.87 $\pm^{ 0.44}_{0.45}$ & -1.87 $\pm^{ 0.13}_{0.13}$ &--& -1.86 $\pm^{ 0.45}_{0.45}$ & -1.87 $\pm^{ 0.13}_{0.13}$ \\
%IRAS 15469-5311 & -2.51 $\pm^{ 1.70}_{1.70}$ & -3.18 $\pm^{ 0.21}_{0.22}$ &--& -2.48 $\pm^{ 1.67}_{1.73}$ & -3.18 $\pm^{ 0.21}_{0.22}$ \\
IRAS 15556-5444 & -1.99 $\pm^{ 0.30}_{0.30}$ & -1.78 $\pm^{ 0.11}_{0.11}$ &--& -1.99 $\pm^{ 0.30}_{0.30}$ & -1.78 $\pm^{ 0.10}_{0.11}$ \\
%NSV 7708 & -2.50 $\pm^{ 1.67}_{1.69}$ & -2.67 $\pm^{ 0.53}_{0.55}$ &--& -2.53 $\pm^{ 1.72}_{1.65}$ & -2.66 $\pm^{ 0.53}_{0.55}$ \\
IRAS 17038-4815 & -1.95 $\pm^{ 0.38}_{0.36}$ & -1.66 $\pm^{ 0.11}_{0.11}$ &--& -1.94 $\pm^{ 0.37}_{0.36}$ & -1.65 $\pm^{ 0.12}_{0.12}$ \\
LR Sco & -0.71 $\pm^{ 0.42}_{0.49}$ & -1.95 $\pm^{ 0.16}_{0.16}$ &--& -0.72 $\pm^{ 0.43}_{0.49}$ & -1.95 $\pm^{ 0.16}_{0.15}$ \\
89 Her & -2.01 $\pm^{ 0.32}_{0.32}$ & -2.04 $\pm^{ 0.31}_{0.31}$ & -2.31 $\pm^{ 0.23}_{0.23}$ & -2.45 $\pm^{ 0.08}_{0.08}$ & -2.45 $\pm^{ 0.08}_{0.08}$ \\
AI Sco & -2.18 $\pm^{ 0.39}_{0.39}$ & -1.83 $\pm^{ 0.12}_{0.12}$ &--& -2.21 $\pm^{ 0.40}_{0.38}$ & -1.84 $\pm^{ 0.12}_{0.12}$ \\
V2053 Oph & -3.63 $\pm^{ 0.89}_{0.85}$ & -2.74 $\pm^{ 0.19}_{0.20}$ &--& -3.71 $\pm^{ 0.96}_{0.83}$ & -2.74 $\pm^{ 0.19}_{0.20}$\\ 
IRAS 18158-3445 & -2.08 $\pm^{ 0.23}_{0.23}$ & -0.03 $\pm^{ 0.02}_{0.04}$ &--& -2.08 $\pm^{ 0.24}_{0.23}$ & -0.03 $\pm^{ 0.02}_{0.04}$ \\
AC Her & -2.05 $\pm^{ 0.26}_{0.26}$ & -2.04 $\pm^{ 0.27}_{0.26}$ & -2.40 $\pm^{ 0.16}_{0.16}$ & -2.32 $\pm^{ 0.06}_{0.06}$ & -2.32 $\pm^{ 0.06}_{0.06}$ \\
%AD Aql & -3.17 $\pm^{ 1.66}_{1.26}$ & -1.78 $\pm^{ 0.29}_{0.32}$ &--& -3.14 $\pm^{ 1.65}_{1.29}$ & -1.80 $\pm^{ 0.29}_{0.33}$ \\
EP Lyr & -1.22 $\pm^{ 0.80}_{1.04}$ & -1.02 $\pm^{ 0.29}_{0.30}$ &--& -1.23 $\pm^{ 0.81}_{1.04}$ & -1.03 $\pm^{ 0.30}_{0.29}$ \\
BD-02 4931 & -1.70 $\pm^{ 0.58}_{0.59}$ & -1.81 $\pm^{ 0.15}_{0.15}$ &--& -1.71 $\pm^{ 0.59}_{0.58}$ & -1.80 $\pm^{ 0.15}_{0.15}$ \\
QY Sge & -1.85 $\pm^{ 0.43}_{0.41}$ & -1.85 $\pm^{ 0.42}_{0.43}$ & -2.48 $\pm^{ 1.70}_{1.71}$ & -2.28 $\pm^{ 0.18}_{0.19}$ & -2.29 $\pm^{ 0.19}_{0.19}$ \\
%AR Pup & -1.82 $\pm^{ 0.31}_{0.31}$ & -1.81 $\pm^{ 0.31}_{0.31}$ & -2.44 $\pm^{ 1.66}_{1.71}$ & -2.25 $\pm^{ 0.10}_{0.10}$ & -2.25 $\pm^{ 0.10}_{0.10}$ \\
%AR Pup2 & -1.98 $\pm^{ 0.78}_{0.77}$ & -2.00 $\pm^{ 0.76}_{0.78}$ & -2.43 $\pm^{ 1.67}_{1.75}$ & -2.21 $\pm^{ 0.17}_{0.17}$ & -2.21 $\pm^{ 0.18}_{0.18}$ \\
%AR Pup3
AR Pup & -1.89 $\pm^{ 0.50}_{0.50}$ & -1.90 $\pm^{ 0.50}_{0.50}$ & -2.53 $\pm^{ 1.73}_{1.67}$ & -2.23 $\pm^{ 0.10}_{0.10}$ & -2.22 $\pm^{ 0.10}_{0.10}$ \\
RV Tau & -1.94 $\pm^{ 0.39}_{0.39}$ & -1.75 $\pm^{ 0.12}_{0.12}$ & -1.90 $\pm^{ 0.14}_{0.15}$ & -2.00 $\pm^{ 0.06}_{0.06}$ & -1.92 $\pm^{ 0.04}_{0.04}$ \\
%RV Tau2 & -1.86 $\pm^{ 0.27}_{0.27}$ & -1.79 $\pm^{ 0.05}_{0.05}$ & -1.90 $\pm^{ 0.15}_{0.15}$ & -2.03 $\pm^{ 0.05}_{0.05}$ & -1.93 $\pm^{ 0.02}_{0.02}$ \\
AY Lep & -3.64 $\pm^{ 0.99}_{0.89}$ & -2.30 $\pm^{ 0.19}_{0.21}$ & -2.55 $\pm^{ 1.72}_{1.68}$ & -3.70 $\pm^{ 0.91}_{0.84}$ & -2.35 $\pm^{ 0.17}_{0.20}$ \\
PS Gem & -1.67 $\pm^{ 0.98}_{1.11}$ & -1.60 $\pm^{ 0.25}_{0.25}$ & -2.55 $\pm^{ 1.69}_{1.67}$ & -2.75 $\pm^{ 0.55}_{0.74}$ & -1.91 $\pm^{ 0.13}_{0.16}$ \\
IRAS 17233-4330 & -1.20 $\pm^{ 0.73}_{0.86}$ & -1.41 $\pm^{ 0.18}_{0.18}$ &--& -1.21 $\pm^{ 0.73}_{0.85}$ & -1.41 $\pm^{ 0.18}_{0.18}$ \\
BD+03 3950 & -2.07 $\pm^{ 0.66}_{0.68}$ & -2.12 $\pm^{ 0.10}_{0.10}$ &--& -2.08 $\pm^{ 0.67}_{0.66}$ & -2.12 $\pm^{ 0.10}_{0.10}$ \\
IW Car & -1.95 $\pm^{ 0.10}_{0.10}$ & -1.94 $\pm^{ 0.11}_{0.11}$ & -2.48 $\pm^{ 1.70}_{1.71}$ & -1.98 $\pm^{ 0.10}_{0.09}$ & -1.99 $\pm^{ 0.10}_{0.09}$ \\
Red Rectangle &--&--& -2.67 $\pm^{ 0.29}_{0.28}$ & -2.66 $\pm^{ 0.29}_{0.30}$ & -2.66 $\pm^{ 0.29}_{0.30}$ \\
\hline
\multicolumn{5}{l}{Notes:} \\
\multicolumn{5}{l}{$^{\ast}$ SPIRE wavelengths are contaminated with background emission}
%\multicolumn{5}{l}{Refs: }
\\
\hline
    \end{tabular}
    
    \label{tab:alphas}
\end{table*}

\begin{table*}%[]
    \caption{Predicted sub-mm fluxes (mJy) for sources detected in two or more bands, based on MCMC samples.}
    \centering
    %\rotatebox{90}{
    \renewcommand{\arraystretch}{1.35}
    \begin{tabular}{lccccc}\hline %ccccc}\hline
    Source & $\alpha$ & 450 & 850 & 1.3\,mm & 3\,mm  \\ \hline\hline
    %v%\bigskip
%\rule{0pt}{3ex}
%\medskip
TW Cam & -2.50 $\pm^{0.32}_{0.33}$ & 33$\pm^{5}_{5}$& 6.8$\pm^{2.5}_{1.9}$ & 2.4$\pm^{1.3}_{0.9}$ & 0.29 $\pm^{0.29}_{0.15}$ \\
DY Ori & -3.30 $\pm^{ 0.33}_{0.43}$ & 36$\pm^{7}_{6}$& 4.4$\pm^{2.1}_{1.6}$ & 1.1$\pm^{0.8}_{0.5}$ & 0.068 $\pm^{0.09}_{0.04}$ \\
CT Ori & -2.46 $\pm^{ 1.66}_{1.72}$ & 12$\pm^{21}_{8}$& 2.4$\pm^{16.4}_{2.2}$ & 0.9$\pm^{12.6}_{0.8}$ & 0.1 $\pm^{6.8}_{0.1}$ \\
SU Gem & -2.19 $\pm^{ 0.61}_{0.62}$ & 46$\pm^{16}_{12}$& 11$\pm^{11}_{6}$ & 4.5$\pm^{6.9}_{2.8}$ & 0.72 $\pm^{2.29}_{0.56}$ \\
%UY CMa & -2.48695891554 $\pm^{ 1.68394572596}_{1.69678727092}$ & 0.0$\pm^{4.88263163439e-155}_{0.0}$& 0.0$\pm^{1.46918128868e-155}_{0.0}$ & 0.0$\pm^{5.18557368513e-156}_{0.0}$ & 0.0 $\pm^{4.05657644692e-157}_{0.0}$ \\
%UY CMa2 
UY CMa& -3.10 $\pm^{1.39}_{1.28}$ & 3.7$\pm^{5.0}_{2.3}$& 0.52$\pm^{2.20}_{0.41}$ & 0.14$\pm^{1.15}_{0.12}$ & 0.010 $\pm^{0.291}_{0.009}$ \\
%V382 Aur & -2.48122558626 $\pm^{ 1.67784655668}_{1.71622024257}$ & 0.0$\pm^{3.36848496134e-171}_{0.0}$& 0.0$\pm^{8.7754797931e-172}_{0.0}$ & 0.0$\pm^{2.55876165559e-172}_{0.0}$ & 0.0 $\pm^{3.48363491158e-173}_{0.0}$ \\
ST Pup & -2.82 $\pm^{ 1.77}_{1.50}$ & 5.5$\pm^{9.5}_{3.5}$& 0.9$\pm^{6.2}_{0.8}$ & 0.27$\pm^{4.28}_{0.25}$ & 0.026 $\pm^{1.866}_{0.025}$ \\
V421 CMa & -3.14 $\pm^{ 1.72}_{1.32}$ & 4.5$\pm^{5.9}_{2.6}$& 0.59$\pm^{3.32}_{0.47}$ & 0.16$\pm^{1.95}_{0.14}$ & 0.011 $\pm^{0.620}_{0.011}$ \\
U Mon & -1.98 $\pm^{ 0.03}_{0.04}$ & 640$\pm^{27}_{25}$& 182$\pm^{5}_{5}$ & 79$\pm^{2}_{2}$ & 15.1 $\pm^{0.7}_{0.6}$ \\
V390 Vel & -2.84 $\pm^{ 0.25}_{0.25}$ & 833$\pm^{85}_{78}$& 137$\pm^{37}_{29}$ & 41$\pm^{17}_{12}$ & 3.8 $\pm^{2.8}_{1.6}$ \\
%BZ Pyx & -2.49578457865 $\pm^{ 1.69555583932}_{1.70472425585}$ & 0.0$\pm^{2.48115491505e-153}_{0.0}$& 0.0$\pm^{4.1005411476e-154}_{0.0}$ & 0.0$\pm^{1.3859570949e-154}_{0.0}$ & 0.0 $\pm^{1.70197044532e-155}_{0.0}$ \\
%IRAS 09144-4933 & -2.48095666627 $\pm^{ 1.7060379051}_{1.70298460098}$ & 0.0$\pm^{5.16457633439e-168}_{0.0}$& 0.0$\pm^{7.73036692137e-169}_{0.0}$ & 0.0$\pm^{2.1251183819e-169}_{0.0}$ & 0.0 $\pm^{5.16753334039e-170}_{0.0}$ \\
IRAS 09400-4733 & -3.11 $\pm^{ 0.62}_{0.71}$ & 18$\pm^{7}_{6}$& 2.6$\pm^{2.6}_{1.4}$ & 0.68$\pm^{1.11}_{0.45}$ & 0.051 $\pm^{0.174}_{0.041}$ \\
%GP Cha & -2.42444708052 $\pm^{ 1.66744813751}_{1.7521966195}$ & 0.0$\pm^{4.94613513888e-159}_{0.0}$& 0.0$\pm^{1.11747439634e-159}_{0.0}$ & 0.0$\pm^{4.26738153217e-160}_{0.0}$ & 0.0 $\pm^{4.72538860687e-161}_{0.0}$ \\
AG Ant & -1.13 $\pm^{ 0.48}_{0.49}$ & 40$\pm^{10}_{8}$& 20$\pm^{12}_{7}$ & 12$\pm^{11}_{6}$ & 4.7 $\pm^{8.8}_{3.1}$ \\
HR 4226 & -2.03 $\pm^{ 0.25}_{0.25}$ & 655$\pm^{68}_{62}$& 180$\pm^{49}_{38}$ & 76$\pm^{31}_{22}$ & 14 $\pm^{10}_{6}$ \\
%V802 Car & -2.61637434363 $\pm^{ 1.74773035182}_{1.66108737785}$ & 0.0$\pm^{2.54800216288e-165}_{0.0}$& 0.0$\pm^{6.13612037924e-166}_{0.0}$ & 0.0$\pm^{2.30285634503e-166}_{0.0}$ & 0.0 $\pm^{3.71291917921e-167}_{0.0}$ \\
%GK Car & -2.42252474301 $\pm^{ 1.6619553812}_{1.71239126087}$ & 0.0$\pm^{1.01820259979e-160}_{0.0}$& 0.0$\pm^{2.16684255176e-161}_{0.0}$ & 0.0$\pm^{7.7242382937e-162}_{0.0}$ & 0.0 $\pm^{1.0852251246e-162}_{0.0}$ \\
AF Crt & -2.42 $\pm^{ 1.38}_{1.50}$ & 8.4$\pm^{7.6}_{4.4}$& 1.8$\pm^{5.7}_{1.5}$ & 0.65$\pm^{4.02}_{0.58}$ & 0.086 $\pm^{1.865}_{0.083}$ \\
RU Cen & -1.97 $\pm^{ 0.31}_{0.31}$ & 188$\pm^{26}_{23}$& 54$\pm^{20}_{14}$ & 23$\pm^{13}_{8}$ & 4.5 $\pm^{4.5}_{2.2}$ \\
SX Cen & -2.87 $\pm^{ 0.83}_{0.97}$ & 17$\pm^{8}_{6}$& 2.7$\pm^{3.9}_{1.8}$ & 0.80$\pm^{1.94}_{0.62}$ & 0.073 $\pm^{0.42}_{0.066}$ \\
V1123 Cen & -2.50 $\pm^{ 0.66}_{0.66}$ & 67$\pm^{26}_{19}$& 14$\pm^{15}_{7}$ & 4.7$\pm^{8.3}_{3.0}$ & 0.59 $\pm^{2.20}_{0.46}$ \\
%V1123 Cen2 & -2.36 $\pm^{ 0.98}_{1.00}$ & 74$\pm^{54}_{32}$& 16$\pm^{36}_{11}$ & 6.0$\pm^{23}_{4.8}$ & 0.83 $\pm^{8.34}_{0.76}$ \\
IRAS 13258-8103 & -3.91 $\pm^{ 0.86}_{0.73}$ & 14$\pm^{9}_{5}$& 1.2$\pm^{2.2}_{0.7}$ & 0.23$\pm^{0.69}_{0.16}$ & 0.086 $\pm^{0.062}_{0.072}$ \\
EN TrA & -1.86 $\pm^{ 0.45}_{0.45}$ & 88$\pm^{20}_{16}$& 27$\pm^{17}_{10}$ & 12$\pm^{12}_{6}$ & 2.6 $\pm^{4.8}_{1.7}$ \\
%IRAS 15469-5311 & -2.47723716098 $\pm^{ 1.67371879425}_{1.72953296001}$ & 0.0$\pm^{7.65831692405e-163}_{0.0}$& 0.0$\pm^{1.63526248916e-163}_{0.0}$ & 0.0$\pm^{5.73492950923e-164}_{0.0}$ & 0.0 $\pm^{9.48661538739e-165}_{0.0}$ \\
IRAS 15556-5444 & -1.99 $\pm^{ 0.30}_{0.30}$ & 201$\pm^{27}_{24}$& 57$\pm^{20}_{15}$ & 24$\pm^{13}_{8}$ & 4.6 $\pm^{4.5}_{2.3}$ \\
%NSV 7708 & -2.52830865527 $\pm^{ 1.7167241488}_{1.64748956778}$ & 0.0$\pm^{1.37911198789e-168}_{0.0}$& 0.0$\pm^{3.24756787196e-169}_{0.0}$ & 0.0$\pm^{1.08140115079e-169}_{0.0}$ & 0.0 $\pm^{1.94957307922e-170}_{0.0}$ \\
IRAS 17038-4815 & -1.94 $\pm^{ 0.37}_{0.36}$ & 149$\pm^{25}_{22}$& 43$\pm^{19}_{13}$ & 19$\pm^{13}_{8}$ & 3.7 $\pm^{4.8}_{2.1}$ \\
LR Sco & -0.72 $\pm^{ 0.43}_{0.49}$ & 52$\pm^{12}_{10}$& 33$\pm^{18}_{13}$ & 24$\pm^{21}_{12}$ & 13 $\pm^{22}_{9}$ \\
89 Her & -2.45 $\pm^{ 0.08}_{0.08}$ & 171$\pm^{11}_{11}$& 36$\pm^{2}_{2}$ & 13$\pm^{1}_{1}$ & 1.6 $\pm^{0.2}_{0.2}$ \\
AI Sco & -2.21 $\pm^{ 0.40}_{0.38}$ & 82$\pm^{16}_{13}$& 20$\pm^{10}_{7}$ & 7.9$\pm^{6.2}_{3.4}$ & 1.2 $\pm^{1.8}_{0.7}$ \\
V2053 Oph & -3.71 $\pm^{ 0.96}_{0.83}$ & 12$\pm^{8}_{5}$& 1.2$\pm^{2.3}_{0.7}$ & 0.24$\pm^{0.85}_{0.18}$ & 0.011 $\pm^{0.098}_{0.094}$ \\
IRAS 18158-3445 & -2.08 $\pm^{ 0.24}_{0.23}$ & 392$\pm^{36}_{34}$& 105$\pm^{26}_{21}$ & 43$\pm^{16}_{12}$ & 7.6 $\pm^{5.1}_{3.1}$ \\
AC Her & -2.32 $\pm^{ 0.06}_{0.06}$ & 441$\pm^{25}_{24}$& 100$\pm^{4}_{4}$ & 37$\pm^{2}_{2}$ & 5.3$\pm^{0.5}_{0.5}$ \\
AD Aql & -3.14 $\pm^{ 1.65}_{1.29}$ & 4.9$\pm^{6.1}_{2.7}$& 0.65$\pm^{3.36}_{0.50}$ & 0.17$\pm^{1.92}_{0.15}$ & 0.012 $\pm^{0.581}_{0.012}$ \\
EP Lyr & -1.23$\pm^{0.81}_{1.04}$ & 19$\pm^{11}_{7}$& 8.8$\pm^{11.8}_{5.7}$ & 5.2$\pm^{11.7}_{4.0}$ & 1.9 $\pm^{9.8}_{1.7}$ \\
BD-02 4931 & -1.71 $\pm^{ 0.59}_{0.58}$ & 56$\pm^{18}_{14}$& 19$\pm^{17}_{8.8}$ & 9.2$\pm^{13.0}_{5.4}$ & 2.2 $\pm^{6.5}_{1.6}$ \\
QY Sge & -2.28$\pm^{ 0.18}_{0.19}$ & 102$\pm^{10}_{9}$& 24$\pm^{4.6}_{3.7}$ & 9.0$\pm^{2.5}_{1.9}$ & 1.3 $\pm^{0.6}_{0.4}$ \\
%AR Pup & -2.24947914436 $\pm^{ 0.0957604669344}_{0.0967458712752}$ & 0.474098716807$\pm^{0.0315496596722}_{0.0305712455687}$& 0.113262605381$\pm^{0.00973467427791}_{0.00899657010178}$ & 0.0435577316035$\pm^{0.00512318308817}_{0.00470313204331}$ & 0.00663235747278 $\pm^{0.00133962913439}_{0.00112364401156}$ \\
%AR Pup2 & -2.20567296464 $\pm^{ 0.1748966031}_{0.174820327057}$ & 0.432142967471$\pm^{0.0801683724191}_{0.0663816177624}$& 0.106339910438$\pm^{0.0120233144858}_{0.0104190405745}$ & 0.0417574426517$\pm^{0.00502723218333}_{0.0045665146805}$ & 0.00659675546504 $\pm^{0.00164708324975}_{0.00131135074002}$ \\
%AR Pup3 
AR Pup & -2.23 $\pm^{ 0.10}_{0.10}$ & 449$\pm^{37}_{35}$& 109$\pm^{10}_{9}$ & 42$\pm^{5}_{5}$ & 6.6 $\pm^{1.3}_{1.1}$ \\
RV Tau & -2.00 $\pm^{ 0.06}_{0.06}$ & 156$\pm^{11}_{10}$& 44$\pm^{3}_{2}$ & 19$\pm^{1}_{1}$ & 3.5$\pm^{0.4}_{0.3}$ \\
%RV Tau2 & -2.02906723312 $\pm^{ 0.0485414051961}_{0.0491959461972}$ & 0.164740684048$\pm^{0.00755534696824}_{0.00729099249888}$& 0.0453037679619$\pm^{0.00232758299053}_{0.00218398945724}$ & 0.019130888323$\pm^{0.00122765520487}_{0.00115528581612}$ & 0.00350601755471 $\pm^{0.000350273871383}_{0.000321377363198}$ \\
AY Lep & -3.70$\pm^{ 0.91}_{0.84}$ & 12$\pm^{8}_{5}$& 1.1$\pm^{2.1}_{0.7}$ & 0.23$\pm^{0.73}_{0.17}$ & 0.010$\pm^{0.082}_{0.009}$ \\
PS Gem & -2.75 $\pm^{ 0.55}_{0.74}$ & 21$\pm^{8}_{6}$& 3.8$\pm^{2.5}_{2.0}$ & 1.2$\pm^{1.2}_{0.8}$ & 0.12$\pm^{0.25}_{0.09}$ \\
IRAS 17233-4330 & -1.21 $\pm^{ 0.73}_{0.85}$ & 141$\pm^{62}_{46}$& 66$\pm^{76}_{39}$ & 39$\pm^{76}_{28}$ & 14 $\pm^{63}_{12}$ \\
BD+03 3950 & -2.08 $\pm^{ 0.67}_{0.66}$ & 91$\pm^{39}_{27}$& 24$\pm^{28}_{13}$ & 10$\pm^{19}_{6}$ & 1.8 $\pm^{7.1}_{1.4}$ \\
IW Car & -1.98$\pm^{ 0.10}_{0.09}$ & 841$\pm^{42}_{39}$& 238$\pm^{27}_{24}$ & 102$\pm^{16}_{14}$ & 19$\pm^{5}_{4}$ \\
Red Rectangle & -2.66$\pm^{0.29}_{0.30}$ & 34$\pm^{5}_{4}$& 6.3$\pm^{0.9}_{0.8}$ & 2.0$\pm^{0.5}_{0.4}$ & 0.22 $\pm^{0.13}_{0.08}$ \\%\rule{0pt}{3ex}
    \hline
    \end{tabular}
    %}
    \label{tab:predictions}
\end{table*}

\begin{figure}
    \centering
    \includegraphics[width=0.45\textwidth]{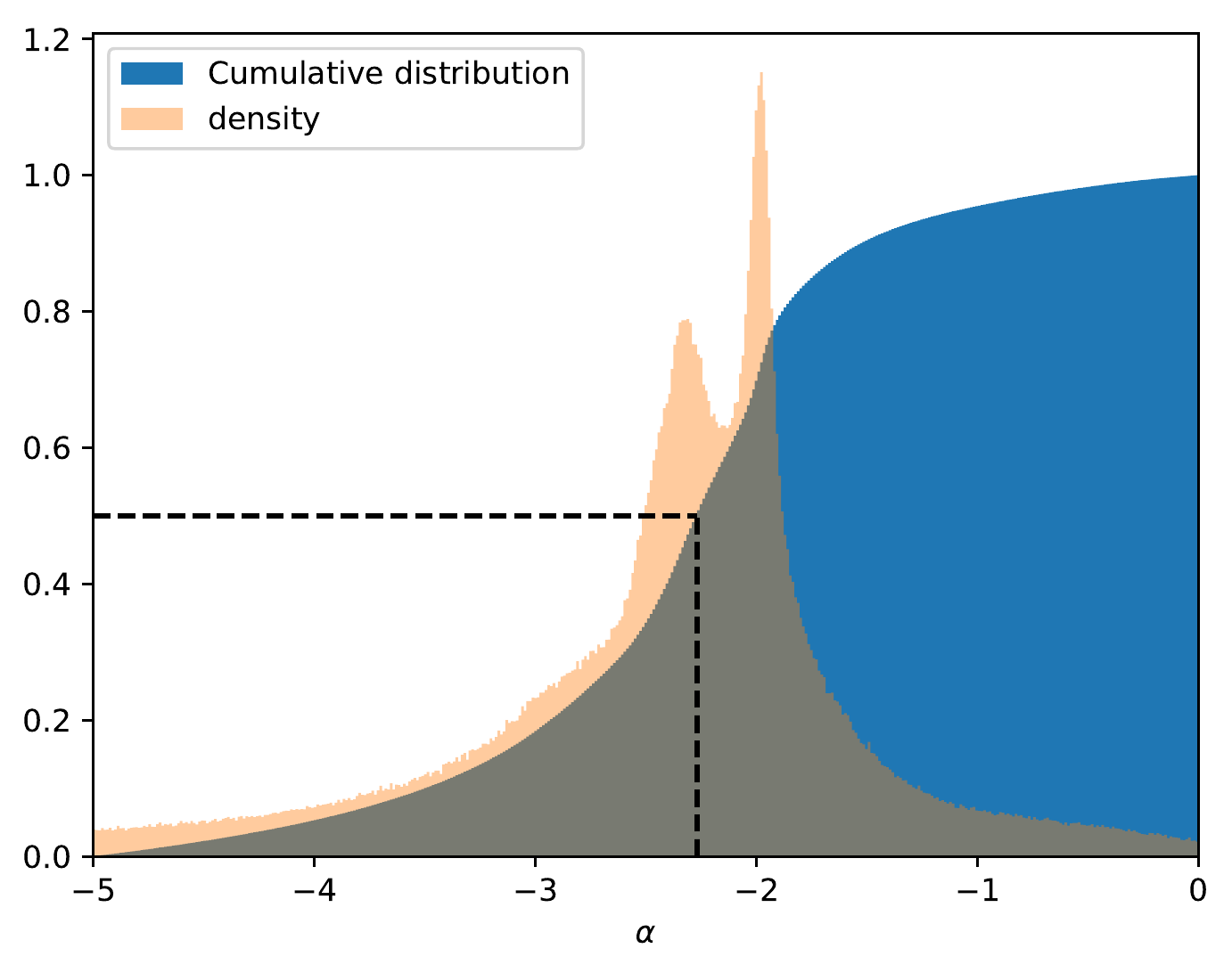}
    \caption{Histogram of the distribution of $\alpha$ values from MCMC realisations, %{\it Left}: 
    produced by merging the full set of MCMC samples after burn in for all sources with detections in two or more bands, %. {\it Right}: A
    after removing AG Ant, AF Crt and EP Lyr, which have poor quality fits. Including these sources would increase the tails of the distribution without affecting the core of the density. The dashed line indicates the median of the distribution.}
    \label{fig:alphahists}
\end{figure}

Using the full set of MCMC realisations (after an appropriate burn-in period), we construct an estimate of the distribution of spectral indices. % using Kernel Density Estimation (KDE). % for each subset of the data.
%Several kernels were employed to explore the importance of this choice.
%The results are shown in Fig.~\ref{}.
%{ PS: The large dataset makes KDE very expensive. For the moment, these are replaced with histograms. }
The sharp peak at $-2$ results from the group of sources with long-wavelength data and very well constrained spectral indices, and the broader bump describes the wider population, with long tails toward very steep and very shallow slopes.
A second peak exists at $\sim-2.3$, however it is unclear whether this reflects a genuine separation into two groups.

%This includes fitting the power law to the data and producing all plots presented in this section and in the Discussion below.
%\subsection{Disc masses?}
%{ PS: Should we try doing this? Do we have enough reliable distance estimates to make fitting a MBB a realistic option?}

\section{Discussion}
\label{sect:discussion}
The dataset shows a clear abundance of flat emission across the FIR--sub-mm range.
The distributions of $\alpha$ are clearly dominated by values close to 2, regardless of the choice of dataset.
However, there are clear differences that depend on the choice of wavelengths used to derive $\alpha$.

\subsection{Importance of long-wavelength photometry}
\begin{figure}
    \centering
    \includegraphics[width=0.45\textwidth]{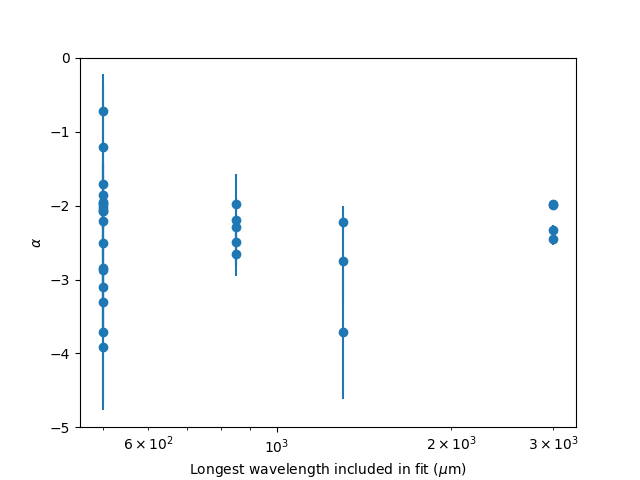} \\
    \includegraphics[width=0.45\textwidth]{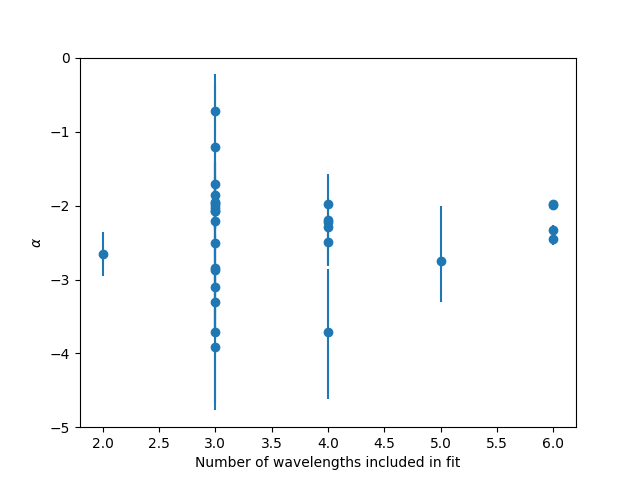}
    \caption{Fitted spectral index as a function of the longest wavelength included in the fit.}
    \label{fig:wavealpha}
\end{figure}
%- Plot of wavelengths
The choice of wavelength range over which to fit the spectral index has a significant effect on the outcome of the fit. 
Figure~\ref{fig:wavealpha} shows the fitted value of $\alpha$ as a function of the longest wavelength included in the fit.
There is a clear reduction in the uncertainty on the spectral index as longer wavelengths are used.
This is in part due to the increased number of points and partly a result of the larger separation in wavelength between the longest and shortest wavelengths.
However, there is no real trend in the value of the spectral index.

The importance of the choice of wavelengths can also be explored by looking at the values of $\alpha$ derived for the same object with different wavelengths.
In particular, the 2nd and 4th columns of Tab.~\ref{tab:alphas} can be compared for those objects with additional data.
In general, the addition of longer-wavelength data tends to result in a steeper slope by up to roughly 25\%.
However, this trend results in a significant change\footnote{defined as $\frac{\left|\Delta\alpha\right|}{\sqrt{\sigma_{\alpha_1} + \sigma_{\alpha_2}}} \geq 1$} only in the two cases of 89 Her and AC Her, which have a plethora of high-precision detections.
There is one exception, TW~Cam, where the slope becomes shallower, possibly because of the lack of 500\,$\mu$m detection, which warrants further investigation.

A further important consideration is whether these changes in spectral index are real or an artefact of biases.
This factor is considered below in Sect~\ref{sec:alphagrow}, where we explore the role of confounding effects in detail.

\subsection{Link between spectral index and grain growth}\label{sec:alphagrow}
Although grain growth does cause shallow spectral indices, can we exclude other mechanisms that produce the same effect? 
These can broadly be divided into three cases:
\begin{enumerate}
    \item different emission mechanisms e.g. free-free emission;
    \item radiative-transfer effects arising from high optical depth or low dust temperatures; or 
    \item changes in dust refractive indices.
\end{enumerate}

The first of these can be easily examined through observations at cm wavelengths.
Where this has been done \citep[e.g.][]{sahai2011} it has been shown that the free-free contribution at wavelengths $\leq 3$\,mm is negligible, and that the 0.3 -- 3\,mm emission is dominated by dust in post-AGB discs.

The role temperature and opacity effects can, to a certain extent, be seen in the change of $\alpha$ as different wavelength ranges are included in the fit.
In particular, the inclusion of PACS observations, which lie close to the peak of the Planck function for the temperatures expected for an optically thick circumstellar disc, biases the fit toward shallower spectral indices.
As explored above, this trend does not continue for longer wavelengths, with the inclusion of millimetre wavelengths primarily reducing the uncertainties on $\alpha$ rather than modifying the best-fitting value. %{ PS: This is an irritating contradiction - when you look at the population of final fits against wavelengths (Fig.~\ref{fig:wavealpha}) there's no evidence for different groups, but when you compare different choices of wavelength for the same object there is a clear difference. I'm not sure how to describe this in terms relevant to this section.}
While this can only be conclusively excluded by combining spatially-resolved multi-wavelength observations with detailed radiative-transfer modelling, the lack of any trend of the calculated $\alpha$ values with wavelength suggests that it does not impact the findings of our analysis in any significant manner.%is not significant.

The final case could arise from changes in dust composition, or perhaps from a dependence of the optical constants on dust temperature \citep[e.g.][]{Mennella1998,boudet2005,zeidler2015}.
This is most difficult to exclude, as it may also introduce some wavelength-dependence of the spectral index.
However, these effects typically introduce a change in the spectral index by at most $\pm$ 1. 
As the distribution of $\alpha$ values is very sharply constrained to the region $-2.5 \leq \alpha \leq -2$, it is clear that only a part of the difference from interstellar dust \citep[expected to have $\alpha\sim-3.6$ based on][]{PlanckCollaboration2014} could come from changes in composition or temperature-dependent opacity.

It is therefore clear that a significant fraction of the sources in our sample have undergone grain growth to sizes sufficient to alter the spectral index, although whether this is the only effect at play requires further obseravations.
Models of the variation of FIR/(sub-)mm dust opacity with grain size suggest that, in order to produce the values we observe, the grains must have grown to sizes at least $\sim 1$\,mm \citep[e.g. Fig. 4 in][]{testi2014} if grain growth alone is responsible.
Even assuming that 50\% of the variation arises due to other effects (i.e. a change in $\alpha$ of $\sim 0.5 - 0.8$ needs to be explained), the dust in these discs must have grown to $\sim 500$\,$\mu$m \citep{testi2014}.

{
 
\subsection{Mid-IR -- far-IR comparison}
}

\begin{figure}
    \centering
    \includegraphics[width=0.47\textwidth]{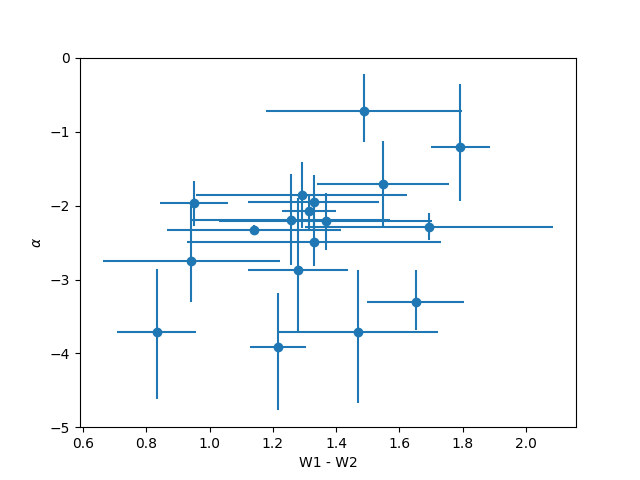} \\
    \includegraphics[width=0.47\textwidth]{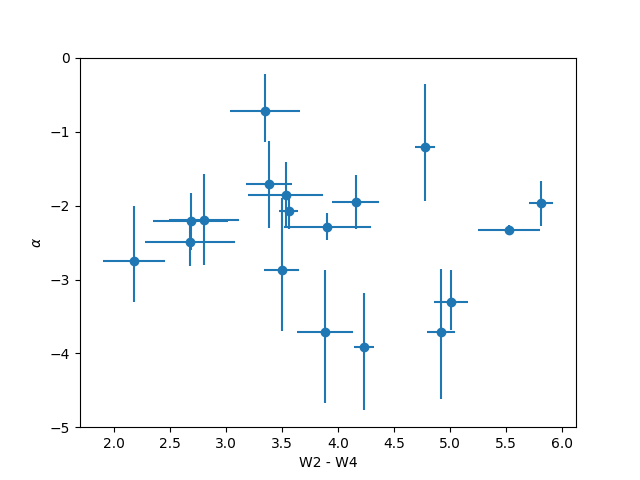}
    \caption{{\it Top}: \textit{WISE} W1-W2 against $\alpha$ {\it Bottom}: W2-W4 against $\alpha$. No correlations between these observables are apparent.}
    \label{fig:WISEComp}
\end{figure}

\citet{deruyter2006} showed that post-AGB disc sources are most similar to the Type II Herbig Ae/Be stars, rather than Type 1 HAeBe or T Tauri stars.
As the distinction between Type I and II is believed to be related to the structure of the discs \citep[e.g.][]{Garufi2017}, we also explore the relationship between the mid-infrared colours, which are expected to reflect the inner-disc structure, and the spectral index in Fig.~\ref{fig:WISEComp}. 
We obtain photometry of our targets at near- and mid-infrared wavelengths from the \textit{Wide-field Infrared Survey Explorer}'s AllWISE catalogue \citep{2010Wright}. \textit{WISE} covered the whole sky in four wavebands (denoted W1 to W4, respectively) with effective wavelengths of 3.4, 4.6, 12 and 22 $\mu$m.
The W1-W2 colour is sensitive to the hottest dust, and hence we expect it to probe how close the inner rim of the disc is to the sublimation radius: the redder the colour, the cooler the dust and hence the further it is from the sublimation radius. 
The W2-W4 colour, on the other hand, reflects the ratio of dust at $\sim$~800\,K and at $\sim$ 200\,K; red colours indicate that the disc hosts a large gap or inner hole.
However, neither plot shows a strong correlation with $\alpha$.
A weak correlation is visible with the W1-W2 colour, but it is possible that this is attributable to a change in the mid-infrared properties of the dust with grain growth (larger grains tend to be cooler) rather than a trend in which discs with larger inner radii have larger grains.
No correlation is apparent between the W1-W2 colour and the orbital period of the binary (where available), indicating that disc-binary interaction does not dominate the location of the warmest dust/disc inner edge.

{
Another way of examining this comparison between grain growth and the mid-IR colours of the discs is to consider the fraction of stellar radiation reprocessed by the discs. 
Using the ratio of $L_{\rm IR} / L_{\ast}$ calculated by \citet{deruyter2006}, we look for a correlation in Fig.~\ref{fig:lir}.
As above, this does not reveal any evidence for a trend, with roughly uniform scatter. We expect that the precision of the alpha determination must reach the ~1\% level (an improvement in precision of roughly two orders of magnitude, corresponding to our most precise alpha measurements with the longest lever-arm in wavelength) before it is useful to explore this relationship, or a factor of 10 more sources must be analysed with alpha at the 10\% level.

\begin{figure}
    \centering
    \includegraphics[width=0.47\textwidth]{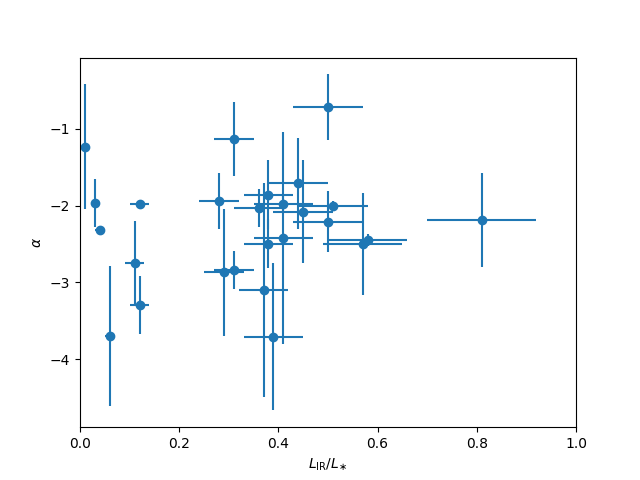}
    \caption{Scatter plot of the fraction of the stellar luminosity reprocessed in the infrared \citep[data from][]{deruyter2006} against $\alpha$. No correlation is apparent, and significantly higher precision on $\alpha$ would be required to probe a relationship.}
    \label{fig:lir}
\end{figure}

It may also be interesting to compare growth from micron to millimetre size; with this in mind we compare our results with those from mid-infrared spectroscopy using the Galactic sample of \citet{2011Gielen}.
However, these methods trace very different regions of the discs in terms of temperature and optical depth; mid-IR spectroscopy is sensitive to $\sim$ micron-sized grains in the upper layers of the warm inner disc, while the FIR/sub-mm spectral index is sensitive to $\sim$ millimetre-sized grains throughout the disc, but particularly the cool midplane.
{  Since these regions are not strongly connected, there is no reason to expect a meaningful correlation between these tracers.}
%Hence, any relationship, or lack thereof, that emerges should be taken with a pinch of salt.

Our sample has 31 sources in common with \citet{2011Gielen}, of which 21 have robust measurements of $\alpha$. 
For these sources, we compare $\alpha$ to the mass-fraction of small, medium and large grains found by \citet{2011Gielen}, by summing the fractions of each for each dust species, to leave the total mass-fraction of each grain size.
Of these, the fraction of small grains can probably be considered the most robust, as a population of grains too large to display features may be hiding in the continuum fit.
These results are shown in Fig.~\ref{fig:giel}.
If grain growth to micron sizes is linked to growth to mm sizes, one would expect to see a correlation between $\alpha$ and the relative abundance of large grains seen in the mid-infrared, and an anticorrelation with the relative abundance of small grains.

\begin{figure*}
    \centering
    \includegraphics[width=\textwidth]{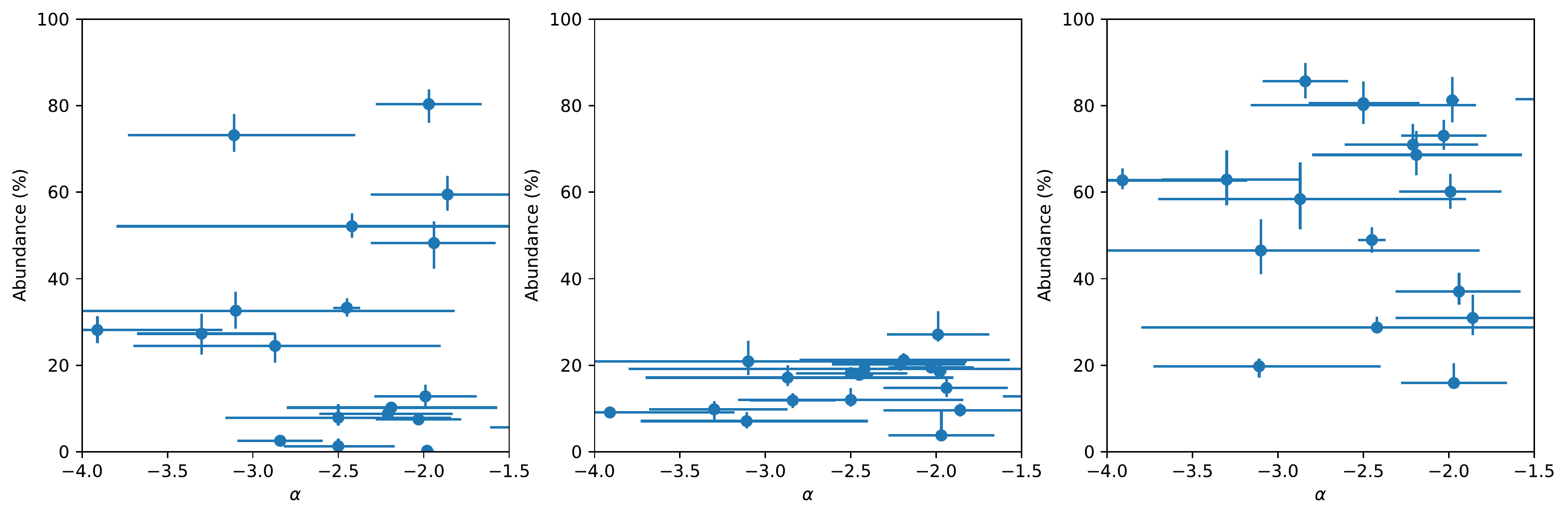}
    \caption{Scatter plots of the mass-fractions of small (left), medium (middle) and large (right) grains against $\alpha$ for the overlap between our sample and that of \citet{2011Gielen}. }
    \label{fig:giel}
\end{figure*}

It may also be interesting to consider the crystalline fraction, as an alternative proxy for dust evolution, and compare this with $\alpha$.
Once again, using the \citet{2011Gielen} sample, we sum the total mass fractions of crystalline Forsterite and Ortho-enstatite, plotting this against $\alpha$ in Fig.~\ref{fig:gielC}.

\begin{figure}
    \centering
    \includegraphics[width=0.47\textwidth]{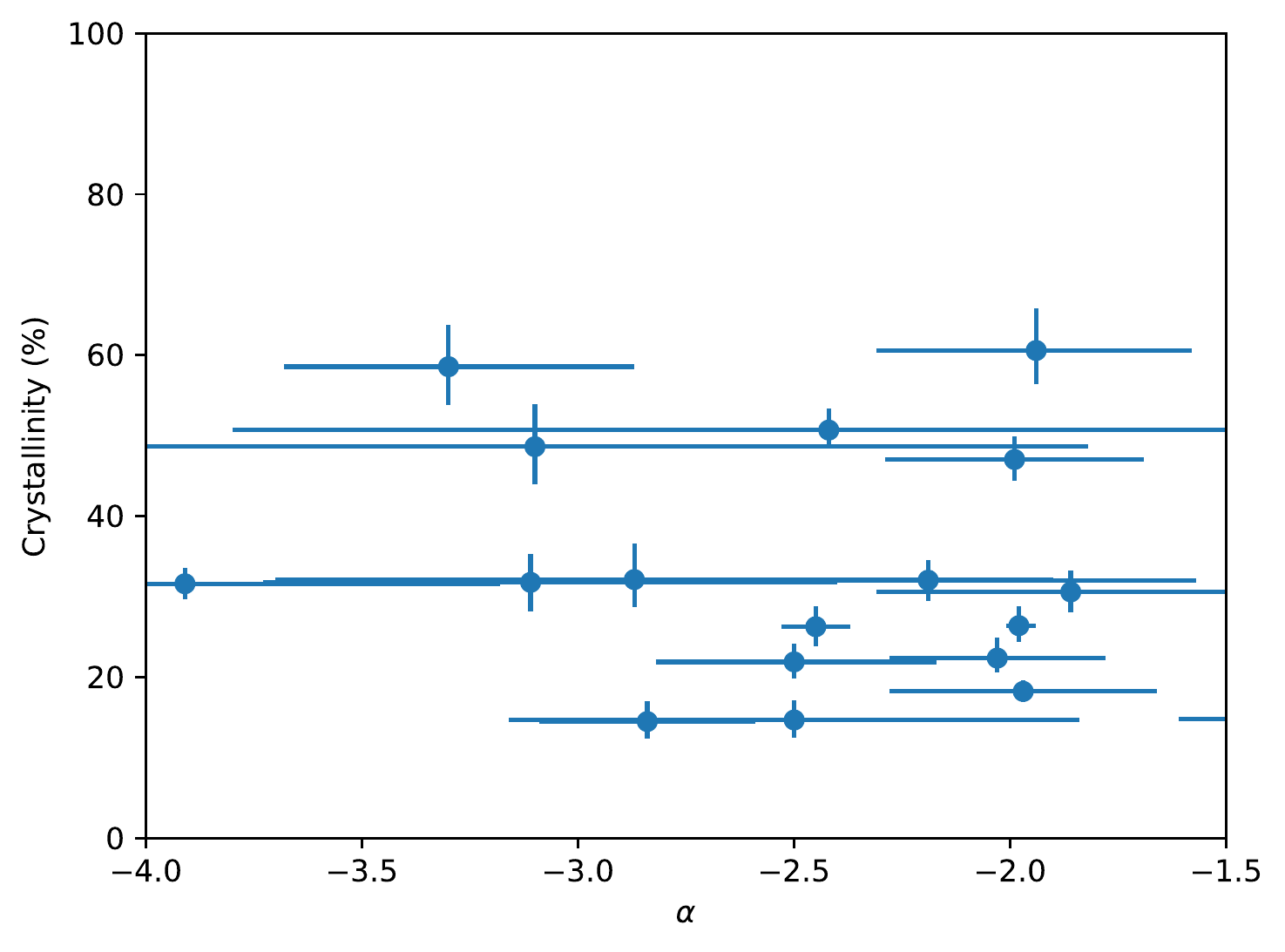}
    \caption{Scatter plot of the crystalline mass-fraction against $\alpha$ for the overlap between our sample and that of \citet{2011Gielen}. }
    \label{fig:gielC}
\end{figure}

%Results go here.
Unsurprisingly, given the weak connection between the regions of the disc that produce these observables, there is no evidence of any trends between grain size as probed by the mid-IR and by the far-IR. 
Similarly, the crystallinity does not correlate with the sub-mm spectral index.
This is most likely a result of the disconnect between the regimes in question, as seen in PPDs \citep[e.g.][]{Ricci2010}.
}

\subsection{Comparison with other types of circumstellar discs}
%How does the distribution of \alpha compare with PPDs and DDs?
Having established that the presence of large dust grains is the most likely cause of the flat FIR/sub-mm spectra, it is interesting to compare this with other types of circumstellar discs.
This may help to elucidate the formation mechanism of the large grains and hence the timescales on which grain-growth may be taking place.

\begin{figure}
    \centering
    \includegraphics[width=0.45\textwidth]{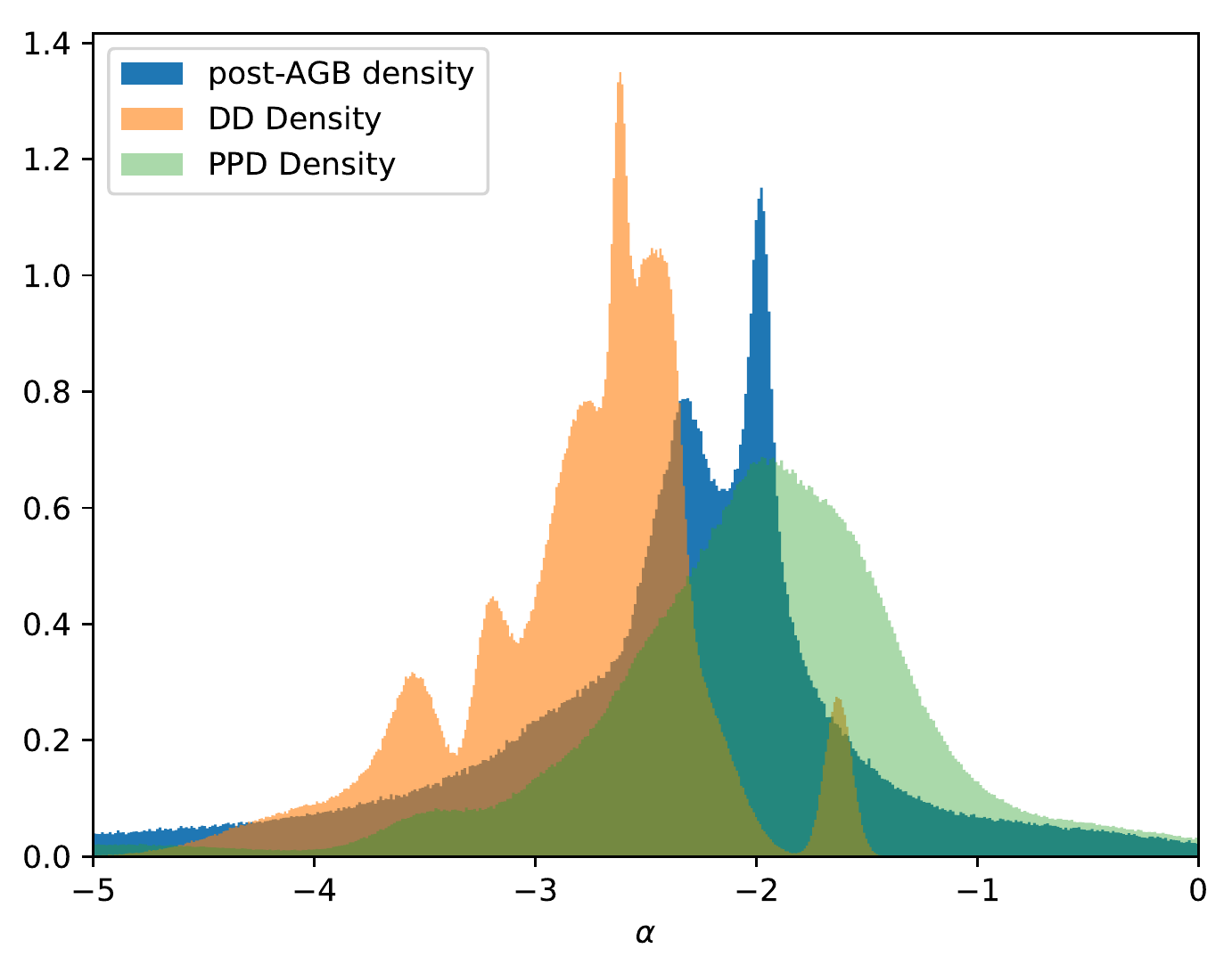}\\
    \includegraphics[width=0.45\textwidth]{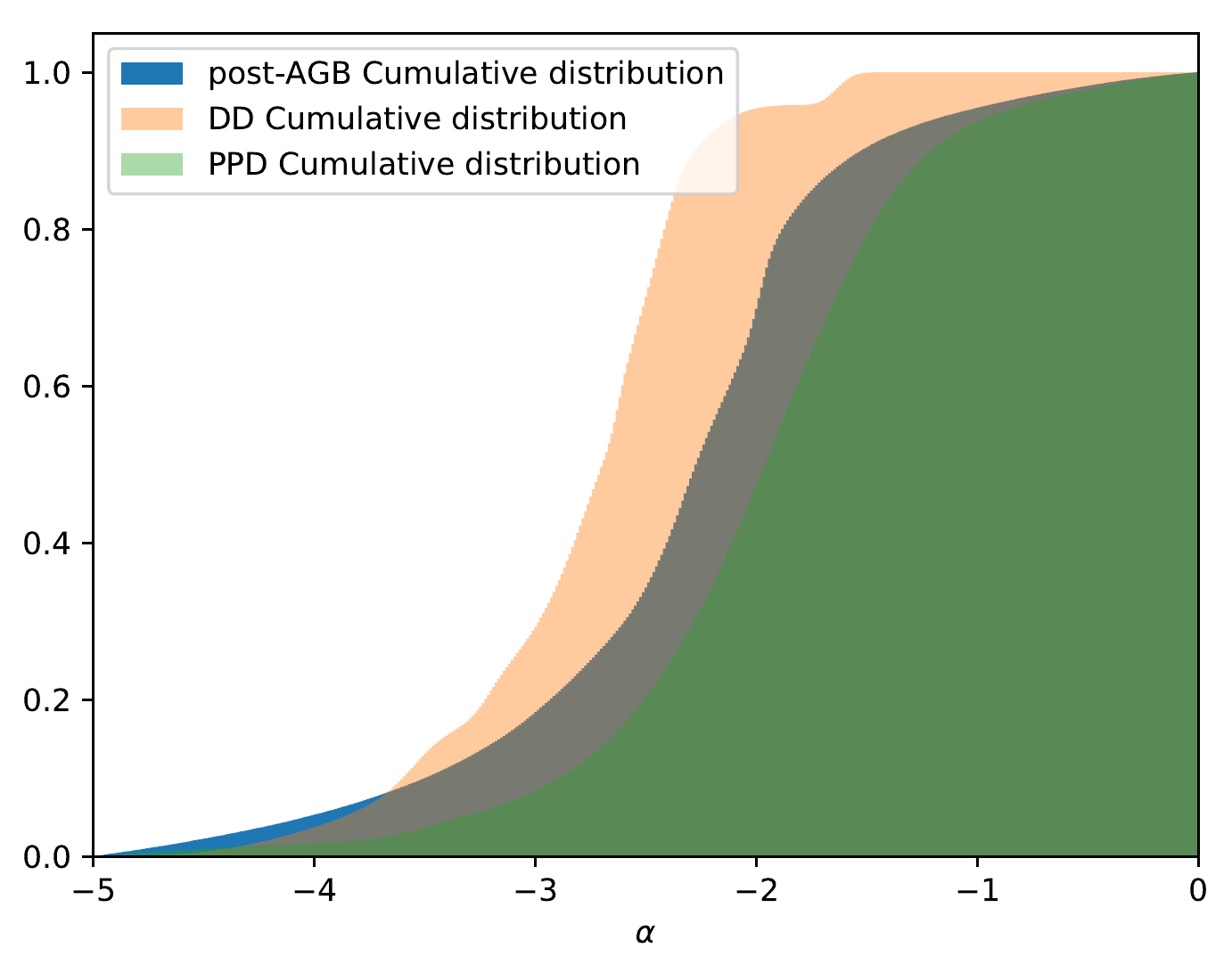}
    \caption{{\it Top}: Comparison of the histograms of the alpha values for our sample (blue) with those for a sample of protoplanetary discs in Taurus \citep[][green]{2005AndrewsWilliams} and a sample of debris discs in the field \citep[][yellow]{2016Macgregor,2017Marshall}. The peak of the distribution for our sample is intermediate to the two samples, but visually more similar to the protoplanetary disc sample. {\it Bottom}: As above, showing the cumulative distributions. It is clear that the distribution of $\alpha$, and hence maximum grain size, is significantly different in debris discs when compared to the other two populations, lacking both very high and very low values of $\alpha$.}
    \label{fig:alpha_comp}
\end{figure}

The two most obvious comparison samples are the gas-rich, optically-thick discs around young pre-main-sequence stars (protoplanetary discs, PPDs), and the gas-poor, optically thin dust discs found around main sequence stars (debris discs, DDs).
In protoplanetary discs, the presence of gas damps the relative velocities of grains, and so collisions result in grain growth, while the lack of gas in debris discs allows a collisional cascade to proceed unimpeded, resulting in the production of a large population of small grains. 
We select two samples: the Taurus protoplanetary disc sample of \citet{2005AndrewsWilliams}, and a sample of field debris discs \citep{2016Macgregor,2017Marshall}.
In order to make a distribution comparable to the histogram shown in Fig.~\ref{fig:alphahists}, we sample from the distributions of published values assuming they are normally distributed with ($\mu, \sigma$) respectively equal to the published value and uncertainty  of alpha. 
We draw samples from these distributions until the number of samples for each group of literature data matches the total number of MCMC samples used to construct the histogram shown in Fig.~\ref{fig:alphahists}.
The results of this sampling are shown in Fig.~\ref{fig:alpha_comp}.

The peak of our sample lies intermediate to the two literature samples, but the { cumulative distribution function} (CDF) is rather more similar to the PPD sample, with a longer tail of very low values of $\alpha$.
We quantify this by examining the sum of the differences between the two other distributions and the post-AGB distribution; this is about 30\% smaller for the PPD than the DD sample.
This supports the expectation that these large grains are indeed formed {\it in situ} from material expelled and trapped in the circumstellar discs, rather than there being two populations of dust; newly-formed small grains, and larger grains produced by a population of parent bodies that somehow survived the post-MS evolution of the star.
Furthermore, the sheer mass of dust would require an unfeasible amount of solid material to be present in the form of planetesimals at the end of the main sequence lifetime of these stars to produce the observed dust through a collisional cascade.
%suggests that, rather than further processing of a population of parent bodies that survived the post-MS evolution of the star, the large dust grains in these systems formed in situ from material expelled and trapped in the circumstellar discs.
%{ PS: It would be good to have dust masses too, to show that they are more comparable to PPDs than anything else - large dust masses imply that the dust must be new and therefore grains must be growing, rather than large dust being liberated as part of the destruction of parent bodies. 
%This means we need distances to the objects to turn fluxes into masses (which will be lower limits because of optical-depth effects), but that could be another labourious job.
%I'll take a look for good resources over the weekend.}

%\subsection{Relationship between stellar and disc properties and spectral index}
%If there is one.

\subsection{Post-AGB timescales}
{ 
The lifetimes of post-AGB stars are dictated by the time it takes for the photosphere to heat up sufficiently to begin ionising the circumstellar envelope, creating a (pre-)planetary nebula.
Single-star models predict that this timescale is a strong function of initial stellar mass, and while it can be as short as decades at high masses it is unlikely to be greater than 10$^{4}$\,yr \citep{2005ARA&A..43..435H,MillerBertolami2016}. 
However, as the accretion of material from the disc tends to inflate the star (leading to a cooler photosphere), this timescale is not likely to be directly applicable in the case of the systems under consideration here, and should be longer. 
Recent work by \citet{Oomen2019} that aims to explain the observed elemental depletions in post-AGBs through accretion shows that the increase in post-AGB lifetime is between a factor of 2 and 5, provided that the accretion rate is high enough (initially $>10^{-7}$\,M$_\odot$\,yr$^{-1}$) and enough mass exists in the disc ($\sim 10^{-2}$\,M$_\odot$) to fuel long-term accretion.
{  Assuming that this can be multiplied by the single-star evolutionary timescale would give a range from 20\,kyr to 50\,kyr.}
Taking this at face value, {  and including an additional uncertainty of a factor of a few,} we would infer that the lifetimes of the post-AGB disc systems, which we expect to come from stars with relatively low initial masses, is typically a few times 10$^{4}$ up to {  perhaps a few times} 10$^{5}$ years.

A small selection of observational constraints also exist for objects in our sample. 
\citet{bujarrabal2007,bujarrabal2015,bujarrabal2016,bujarrabal2017} have resolved the discs of 4 objects in molecular lines (primarily $^{12}$CO and $^{13}$CO), revealing the Keplerian rotation of the discs, and detecting or placing stringent limits on outflows.
Using line radiative-transfer models, they find disc masses of $\sim 10^{-2}$\,M$_\odot$ and, based on the ratio of the disc mass to the outflow, disc lifetimes between 5000 and 20000\,yr, toward the lower end of that expected based on  the work of \citet{Oomen2019}.

However, the uncertainties on these observational timescales are currently poorly known.
The $^{12}$CO lines are typically optically thick, significantly increasing the uncertainty on mass determinations, and while in most cases they have simultaneously fitted $^{13}$CO, with only a single line this is still prone to uncertainty.
In addition, the simplified modelling neglects a number of parameters that have been shown to play important roles in models of protoplanetary discs, such as the structure of the disc.
For example, it has been shown that the amount of flaring in the disc can have a dramatic impact on the line fluxes \citep{Woitke2010}, and if this is not taken into account it will have a large impact on the mass determined for the disc.
Given these limitations, the uncertainties on these models are difficult to quantify but may approach an order of magnitude, bringing them into better agreement with the results of \citet{Oomen2019}.
}

\subsection{Timescales for grain growth} % in post-AGB discs}

{  Our sample of post-AGB stars all have disks with large dust grains, but it is very unlikely that they all have the estimated maximum post-AGB age.  Not all stars will reach that maximum lifetime given a range of initial masses and binary/disk configurations.  Furthermore, the sample likely covers a large range of post-AGB ages between zero and their maximum post-AGB lifetimes.} The existence of a single population {  of grain sizes} implies that either all members started from similar conditions, or they evolve toward similarity on a timescale much shorter than they are evolving\footnote{This is particularly true given that different sources will be evolving on different timescales due to their different masses}.

{  Thus, the grain growth time scale is not constrained in our work by the maximum post-AGB lifetime; instead it must be close to the post-AGB age of the \textit{youngest} star in the sample, at least for that particular system.  This is almost certainly much less than the maximum post-AGB lifetime. Therefore, if we accept the post-AGB lifetime of $10^5$ years discussed in the previous section as an upper limit, the grain-growth timescale must be $<< 10^5$ years and potentially as short as $10^3$ years. }

More specifically, we infer that either dust grains had already grown to large sizes before the post-AGB phase began, or that the timescale for grain growth to mm sizes is effectively instantaneous compared to the post-AGB lifetime i.e., {  $t_{\rm mm} << t_{\rm pAGB}$}, % \citep[i.e., $t_{\rm mm} << 10^{5}$\,yr][]{2005ARA&A..43..435H,MillerBertolami2016,bujarrabal2016,bujarrabal2017}, 
so that all post-AGB discs rapidly converge to the same conditions{  , and hence it follows from the earlier discussion on lifetimes that} $t_{\rm mm} << 10^{\mathbf 5}$\,yr, perhaps by orders of magnitude (i.e. $10^{3} - 10^{4}$\,yr).
The first of these scenarios implies that a significant population of dust became trapped in a long-lived disc while the star was still on the AGB, as the sizes of grains that form in evolved-star winds are typically $\sim 0.05 - 0.5 \mu$m in size depending on the chemistry of the source \citep{2008A&A...491L...1H,2012Natur.484..220N,2015A&A...584L..10S, 2018arXiv181102620N}; these sources are believed to be binaries, and hence this is not implausible. 
%Decin 2019 goes here, L2Pup, R Aqr?
{ Recent observations have shown that some massive, highly-evolved AGB stars do indeed host dense tori \citep{Decin2019} but the high initial masses of these stars makes them unlikely to be the progenitors of the post-AGB disc systems under consideration.}
The majority of a star's mass loss occurs at the end of the AGB in a superwind, meaning that the majority of the material to be trapped should have become so only shortly before the star evolved off the AGB. 
If interactions with the companion are responsible for truncating the AGB phase \citep{2003ARA&A..41..391V} then the mass-loss episode could be even more constrained in time.

In the latter scenario, dust evolution is taking place in these systems on timescales of thousands of years.
This makes them an extreme laboratory for studying the physics of dust growth and processing. 
In comparison, protoplanetary discs are expected to have lifetimes of several megayears.
Models of dust evolution in PPDs are divided: \citet{2010A&A...513A..79B} do not predict the formation of $\gtrsim$100\,$\mu$m particles on timescales $\lesssim10^5$\,yr unless fragmentation is excluded, while \citet{Birnstiel2012} predict the efficient formation of mm-cm size particles in the inner disc on timescales of only 10$^{4}$\,yr. %\citep[e.g.][]{2010A&A...513A..79B} do not predict the formation of $\gtrsim$100\,$\mu$m particles on timescales $\lesssim10^5$\,yr, although the physical conditions in these discs may be considerably different.
However, given the difference in physical conditions between post-AGB and protoplanetary discs, it is not clear how easily this transfers from one to the other.
New dust-evolution models tailored to post-AGB discs may help improve our understanding of grain growth by providing complementary constraints.

As the dust processing in molecular clouds and Class 0/Class 1 YSOs is poorly understood, the initial sizes of dust in PPDs is poorly known, but may be larger than in AGB outflows. 
{ Nevertheless}, the short grain-growth timescale suggested by post-AGB discs challenge models of grain growth in discs in a similar manner to recent observations of polarised (sub-)mm observations of YSOs, which are sometimes interpreted as a result of strong self-scattering and hence the presence of $\sim$mm size grains much earlier, in the Class 0/1 stage of pre-main-sequence evolution \citep[e.g.][]{2015ApJ...814L..28C,2018ApJ...859..165S}.
Similarly, radiative-transfer modelling has suggested the presence of mm or even cm-size grains in a sample of class I discs in Taurus \citep{Sheehan2017}. 
If these short timescales are replicated in PPDs, it may explain the observational evidence of planet formation at the earlier Class 0/1 stages, as seen in e.g. HL Tau \citep{2015ApJ...808L...3A} .

%\subsection{Relationship between disc structure and spectral index}

%{ PS: This depends on getting reliable SpTs from somewhere. SIMBAD isn't a particularly good source for these as they are often low quality. Is this going to involve a labourious literature search?}

%\textbf{JPM:T$_{eff}$ and $\log g$ would be better (as observables) than SpT, but a quick check on SIMBAD/VIZIER suggests that compiling the information required is non-trivial/laborious - anyone got a PhD/Masters student handy?}

%\subsection{Anything else to include?}
%E.g. comparison with other types of objects?
%How do these distributions compare with PPDs? Should we also be comparing with DDs? PPN?

\section{Conclusions}
\label{sect:conclusions}
We present a consistent analysis of \textit{Herschel}, SMA and literature FIR/(sub-)mm fluxes of a sample of 46 post-AGB discs.
By fitting power laws to the emission in this wavelength range, we find that the spectral indices $\alpha$ of these objects are concentrated in the range $-2.5 \leq \alpha \leq -2$, indicating that the grain-size distribution in these discs extends to at least several hundred micron.
The inclusion of long wavelength data is shown to be key to reducing the uncertainty in the determination of $\alpha$ for any given source, and hence surveys of these sources at wavelengths longer than 500\,$\mu$m should be considered a priority to understand whether the multiple sharp peaks seen in the distribution of spectral indices are real, or if there is a single population of spectral indices in post-AGBs.

{ We compare our results with mid-infrared measurements of grain growth for a subset of sources.
Tracers of grain processing in the mid-IR (fraction of $\sim$ micron-sized grains, crystalline fraction) do not correlate with $\alpha$.
This probably reflects the fact that the two tracers are sensitive to dust in very different regions of the discs.}

The large dust mass of these discs and similarity of the distribution of spectral indices to those of protoplanetary discs indicates that this difference in spectral slope from interstellar dust is a result of in-situ grain growth. %, rather than the grinding of parent bodies.
%Furthermore, the sheer mass of dust would require an unfeasible amount of solid material to be present in the form of planetesimals at the end of the main sequence lifetime of these stars to produce the observed dust through a collisional cascade.
Based on the expected sizes of newly formed dust grains around AGB stars, it is clear that a significant amount of grain growth has occurred; grain growth to sizes of at least several hundred micron is ubiquitous in post-AGB disc systems.
The relatively short lifetimes of these discs, and the absence of any evolution of the observed $\alpha$ values with stellar effective temperature (a measure of age), shows that {   all the discs must converge to the same state on a timescale much shorter than their post-AGB evolutionary timescale. Hence} grain growth to these sizes is a rapid process ($\tau_{mm} << 10^{5}$\,yr), making them an extreme laboratory for studying dust growth, with important implications for dust in YSOs and PPDs.

Dust growth has been demonstrated up to millimetre sizes through mutual collisions, but the centimetre size barrier is insurmountable due to mutual velocities and low surface adhesion resulting in bouncing rather than sticking between the largest grains \citep{2010Blum,2012bWindmark,2014Testi}. 
However, if larger grains are already present in the disc further growth of small dust grains beyond the centimetre size barrier is a rapid process \citep[$\sim 10^{4}~$yrs;][]{2012aWindmark}. 
This barrier to growth can also be overcome by assuming some `stickier' dust, % e.g. some fraction of H$_{\rm 2}$O/CO$_{\rm 2}$ ice in their compositions.
%We might expect that, at the end of the main sequence, such large dust grains should be present through the existence of a remnant debris disc around the star to `seed' the post-AGB disc and induce the observed, rapid grain growth.
% These 
which may provide pathways toward the formation of a second generation of planetesimals.

To further understand these enigmatic objects and the insights they can provide on the physics of grain growth, several avenues should be explored.
Firstly, if the analogy with PPDs holds, we expect that different size grains will exist in different regions of the disc. 
In this case, the integrated spectral index is not sufficient, and resolved sub-mm imaging of the discs is required to reveal the importance and timescales of dust drift and settling.
However, to understand whether post-AGB discs really are shorter-lived analogues of PPDs, we must understand their origins.
\citet{2015A&A...578A..77K} recently detected an edge-on disc around the nearby AGB star L$_2$ Pup, {  although its relationship to post-AGB discs remains unclear}.
ALMA observations revealed a candidate sub-stellar object \citep{2016A&A...596A..92K} in the disc, which may be responsible for the recent and rapid formation of the disc.
Continued monitoring of L$_2$ Pup, and other {  evolved systems with discs such as BP Psc \citep{deBoer2017}}, will be key to understanding the formation of these discs and whether the physics is indeed similar to PPDs.

\section*{Acknowledgements}
The authors are grateful to Hans Van Winckel for useful comments and discussion on the topic, and to the anonymous referee whose feedback has helped improve this manuscript.
In addition to the software explicitly cited above, this paper made use of Astropy \citep{astropy_paper,astropy_ascl} and Astroquery \citep{astroquery}. %, ??. Scipy/Numpy/matplotlib?

The Sub-Millimeter Array is a joint project between the Smithsonian Astrophysical Observatory and the Academia Sinica Institute of Astronomy and Astrophysics and is funded by the Smithsonian Institution and the Academia Sinica.

\textit{Herschel} is an ESA space observatory with science instruments provided by European-led Principal Investigator consortia and with important participation from NASA. 

This research has made use of the SIMBAD database, operated at CDS, Strasbourg, France \citep{2000Wenger}. 

This research has made use of NASA's Astrophysics Data System. 

This research has been supported by the Ministry of Science and Technology 
of Taiwan under grants MOST104-2628-M-001-004-MY3 and MOST107-2119-M-001-031-MY3, and by Academia Sinica under grant AS-IA-106-M03.
This work benefited from the FEARLESS collaboration (FatE and AfteRLife of Evolved Solar Systems, PI: S. Ertel).

\bibliographystyle{mnras}
\bibliography{biblio.bib} % if your bibtex file is called example.bib

% Alternatively you could enter them by hand, like this:
% This method is tedious and prone to error if you have lots of references
%\begin{thebibliography}{99}
%\bibitem[\protect\citeauthoryear{Author}{2012}]{Author2012}
%Author A.~N., 2013, Journal of Improbable Astronomy, 1, 1
%\bibitem[\protect\citeauthoryear{Others}{2013}]{Others2013}
%Others S., 2012, Journal of Interesting Stuff, 17, 198
%\end{thebibliography}

%%%%%%%%%%%%%%%%%%%%%%%%%%%%%%%%%%%%%%%%%%%%%%%%%%

%%%%%%%%%%%%%%%%% APPENDICES %%%%%%%%%%%%%%%%%%%%%

% \appendix

% \section{Some extra material}

% If you want to present additional material which would interrupt the flow of the main paper,
% it can be placed in an Appendix which appears after the list of references.

%%%%%%%%%%%%%%%%%%%%%%%%%%%%%%%%%%%%%%%%%%%%%%%%%%

% Don't change these lines
\bsp	% typesetting comment
\label{lastpage}
\end{document}